\documentclass[atmosphere,article,submit,moreauthors,pdftex,10pt,a4paper]{Definitions/mdpi}
\usepackage{gensymb}
\usepackage{amssymb}
\usepackage{afterpage}
\usepackage{textcomp}

\firstpage{1} 
\articlenumber{x}
\doinum{10.3390/------}
\pubvolume{xx}
\pubyear{2018}
\copyrightyear{2018}
\externaleditor{Academic Editor: name}
\history{Received: date; Accepted: date; Published: date}
\Title{Experimental Design of a Prescribed Burn Instrumentation}

\Author{Adam K. Kochanski $^{1}$, Aim\'e Fournier  $^{2}$ and Jan Mandel  $^{2,}$*}
\AuthorNames{Adam K. Kochanski, Aim\'e Fournier and Jan Mandel}

\address{%
$^{1}$ \quad Department of Atmospheric Sciences, University of Utah; adam.kochanski@utah.edu\\
$^{2}$ \quad Department of Mathematical and Statistical Sciences, University of Colorado Denver; aime.fournier@ucdenver.edu, jan.mandel@ucdenver.edu}

\corres{Correspondence: jan.mandel@ucdenver.edu; Tel.: +1-303-315-1703}

\abstract{Observational data collected during experiments, such as the planned Fire and Smoke Model Evaluation Experiment (FASMEE), are critical for evaluating and transitioning coupled fire-atmosphere models like WRF-SFIRE and WRF-SFIRE-CHEM into operational use. Historical meteorological data, representing typical weather conditions for the anticipated burn locations and times, have been processed to initialize and run a set of simulations representing the planned experimental burns. Based on an analysis of these numerical simulations, this paper provides recommendations on the experimental setup such as size and duration of the burns, and optimal sensor placement. New techniques are developed to initialize coupled fire-atmosphere simulations with weather conditions typical of the planned burn locations and times. The variation and sensitivity analysis of the simulation design to model parameters performed by repeated Latin Hypercube Sampling is used to assess the locations of the sensors. The simulations provide the locations for the measurements that maximize the expected variation of the sensor outputs with varying the model parameters.}
\keyword{Coupled atmosphere-fire model; parameter identification; typical weather days; prescribed burn planning; plume rise; sensitivity analysis; sensor placement; FASMEE.}

\MSC{86A10, 62K25}
\begin{document}

%%%%%%%%%%%%%%%%%%%%%%%%%%%%%%%%%%%%%%%%%%
%% Sections that are not mandatory are listed as such. The section titles given are for Articles. Review papers and other article types have a more flexible structure. 

%% Only for the journal Gels: Please place the Experimental Section after the Conclusions

%%%%%%%%%%%%%%%%%%%%%%%%%%%%%%%%%%%%%%%%%%
%\setcounter{section}{-1} %% Remove this when starting to work on the template.
%\section{How to Use this Template}

% The template details the sections that can be used in a manuscript. Sections that are not mandatory are listed as such. The section titles given are for Articles. Review papers and other article types have a more flexible structure. For any questions, please contact the editorial office of the journal or support@mdpi.com. For LaTeX related questions please contact Janine Daum at latex-support@mdpi.com.

\section{Introduction}

Recent advancements in fire-atmosphere numerical modeling have increased the number of physical processes integrated into these coupled models. This greater complexity allows for a more comprehensive representation of the coupled interactions and feedbacks between the fire and the atmosphere. However, as a consequence of these advancements, data requirements for model initialization and validation have increased as well. As coupled fire-atmosphere models utilize local flow properties to parameterize fire progression, emissions, and plume rise~\cite[e.g.,][]{Mandel-2014-RAA}, integrated in-situ measurements are needed for model validation and future development.  

Fire and smoke models are being increasingly relied upon for wildland fire decision making and planning. However, many models are used without adequate validation and evaluation due to the lack of suitable data. Accurate estimates of smoke emissions from wildland fires are highly dependent on reliable characterizations of the area burned, pre-burn biomass of vegetation, fuels, and fuel consumption~\cite{Larkin-2009-BSM,Seiler-1980-EGN,Urbanski-2011-WFE,Liu-2013-EWF}.
Smoke dispersion, on the other hand, strongly depends on the large-scale meteorology and plume injection 
height~\cite{Colarco-2004-TSC,Liu-2008-SAQ},
which is affected by the fire geometry and intensity~\cite{Goodrick-2013-MST}
impacted by local (fire-affected) circulations. Therefore, characterizing fire-atmosphere interactions, including wildland fire behavior and plume dynamics, is fundamental for improving estimates of smoke production and dispersion. The proposed observational data to be collected during the Fire and Smoke Model Evaluation Experiment (FASMEE)~\cite{Ottmar-2017-FSM,Liu-2017-FSM} are critical for evaluating and transitioning coupled fire-atmosphere models like WRF-SFIRE and WRF-SFIRE-CHEM~\cite{Mandel-2011-CAF,Mandel-2014-RAA,Kochanski-2016-TIS} into operational use. 

As the resolution of weather forecasting models increases, the fire-atmosphere interactions and smoke plumes (that had to be treated as sub-grid-scale processes requiring a simplified treatment through external parameterizations) become explicitly res/olvable. Thanks to improved computational capabilities, operational applications of coupled-fire atmosphere models e.g., Israeli Matash system based on WRF-SFIRE, become increasingly feasible~\cite{Matash-2017,Mandel-2014-RAA}. The rapid increase in the resolution of numerical weather prediction products in recent years opens new avenues for development in the near-to-moderate future of integrated systems e.g., WRF-SFIRE-CHEM~\cite{Kochanski-2016-TIS}, capable of resolving, in a~fully-coupled way, fire progression, plume rise, smoke dispersion and chemical transformations.

In order to advance current operational modeling capabilities, rigorous testing, evaluation and model improvements are needed~\cite{Martin-2010-SIH,Mallia-2018-OSP}.
This would allow for assessing  model performance, uncertainties and the key sources of these uncertainties that need improvements.

Currently, there are insufficient observational data available to facilitate this work, especially in the context of coupled fire-atmosphere models requiring integrated datasets to provide information about the fuel, energy release, local micrometeorology, plume dynamics, and chemistry. Without such observational data, the accuracy and validity of even the most sophisticated physics-based modeling approaches cannot be fully assessed. Further, known  gaps in scientific understanding, such as relationships between heat release, wind fields, plume dynamics and feedbacks to wildland fire behavior, cannot be addressed. This paper represents an attempt to: (1) identify the most critical observational needs; (2) assist in the planning of actual experimental burns; and (3) help optimize sensor placement during experimental burns. 
Additional details for the methods employed in this work are explained in~\cite{Kochanski-2017-MSF}. 

%%%%%%%%%%%%%%%%%%%%%%%%%%%%%%%%%%%%%%%%%%
\section{Methods}

%This section may be divided by subheadings. It should provide a concise and precise description of the experimental results, their interpretation as well as the experimental conclusions that can be drawn.

%%%%%%%%%%%%%%%%%%%%%%%%%%%%%%%%%%%%%%%%%%
\subsection{Coupled Atmosphere-Fire Simulation}
\label{sec:simulations}

This study utilizes a suite of numerical simulations, performed using the coupled fire-atmosphere model WRF-SFIRE, which was designed to assist with the planning of the experimental phase of FASMEE. These are simulations of experimental burns to be conducted at the Fishlake, North Kaibab and Fort Stewart sites. Since numerical weather prediction models are unable to forecast weather conditions for burns planned years in advance, we opted to run simulations of experimental burns driven by weather conditions typical of the burn sites during the FASMEE burning seasons. The simulations are driven by time-varying, 3-dimensional historical reanalysis data representing the most typical weather conditions that meet burn requirements and account for the possible local diurnal variability in weather conditions. The most typical days are defined by a statistical analysis of historical weather observations from remote automated weather stations closest to the burn sites for days that meet the burn criteria. The statistical methodology used to select the most typical days is described in Section~\ref{sec:typical}. These days serve as proxies for weather conditions driving the simulations of the experimental burns. 

To support simulations of prescribed burns, models need to accommodate for arbitrary ignition lines that are either continuous (drop-torch ignitions) or consist of a discrete succession of drops of incendiary devices, dispensed from an all-terrain vehicle or an aircraft. The incendiary drops are at arbitrary locations and times, and the distances between drops can be smaller or larger than the fire simulation mesh size, which presents additional modeling challenges. WRF-SFIRE simulates continuous drop-torch ignition at steady speeds along straight lines, developed for the simulation of the FireFlux experiment \cite{Kochanski-2013-EWP}. A generalization of this scheme to complex ignition patterns is unwieldy, so, in this project, we have developed a new ignition mechanism by building on the earlier work on perimeter ignitions~\cite{Kochanski-2016-IFP,Mandel-2012-APD}:
instead of modeling the fire spread until the time of the perimeter is reached, perimeter ignition in WRF-SFIRE prescribes the fire arrival time so that an appropriate atmospheric circulation due to the fire forcing can develop. The new scheme allows the user to specify the latest fire arrival time at the nodes (grid points) of the fire simulation mesh as an additional input array. This enables an integration of the natural fire progression simulated by the model with the ignitions specified by the fire arrival time. 

In WRF-SFIRE, the state of the fire is represented by a so-called level set function, defined in the simulation domain. The fire is burning where the level set function is negative, and an ignition at a given location is implemented by making the level set function take a negative value at that location. However, the level set function is only defined by its values on the nodes of the fire simulation mesh, which are typically spaced several meters to tens of meters apart, while the incendiary device drops are at arbitrary locations. To achieve a sub-fire-mesh representation of the ignition process, the code takes the minimum of the level set function and cones (pointing down) with vertices at the individual ignition points. This approach allows the ignition locations and times not to coincide with the mesh points or the time steps in the model. The aperture of the cones is given by the imposed spread rate in the early stages of the ignition before the fire reaches the mesh node.

\begin{figure}[tbp]
\centering
\includegraphics[width=12cm]{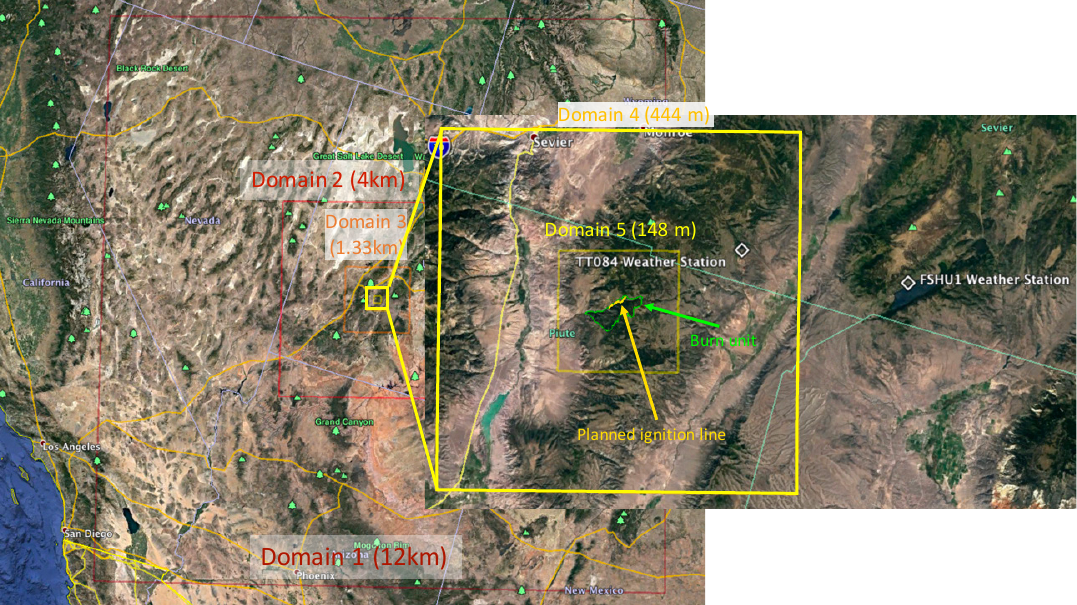}\\
(a)\\\ \\
\includegraphics[width=12cm]{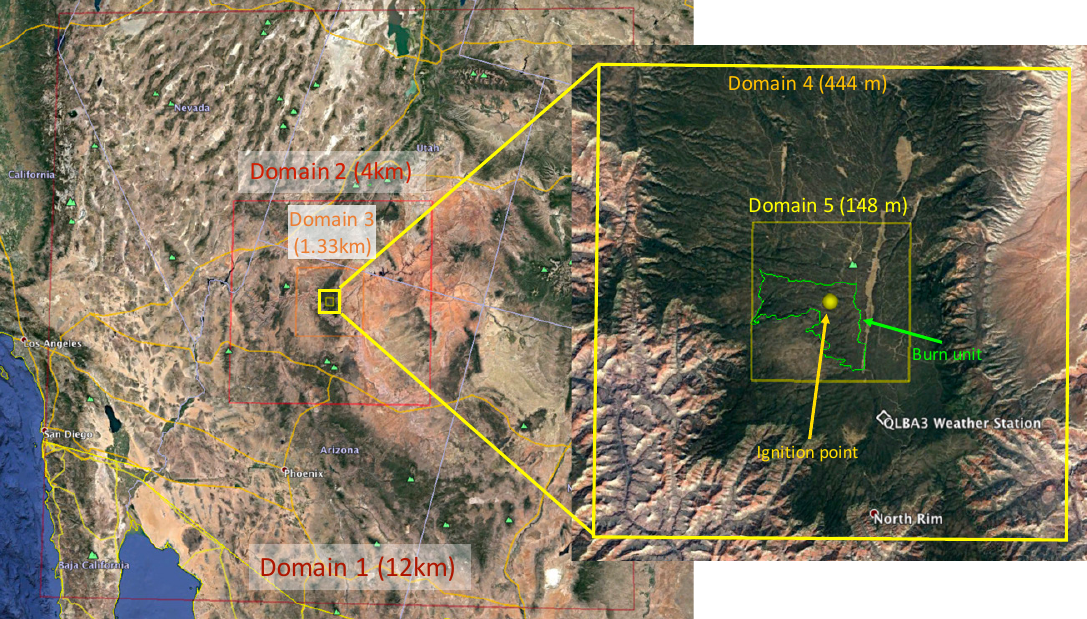}\\
(b)\\\ \\
\includegraphics[width=12cm]{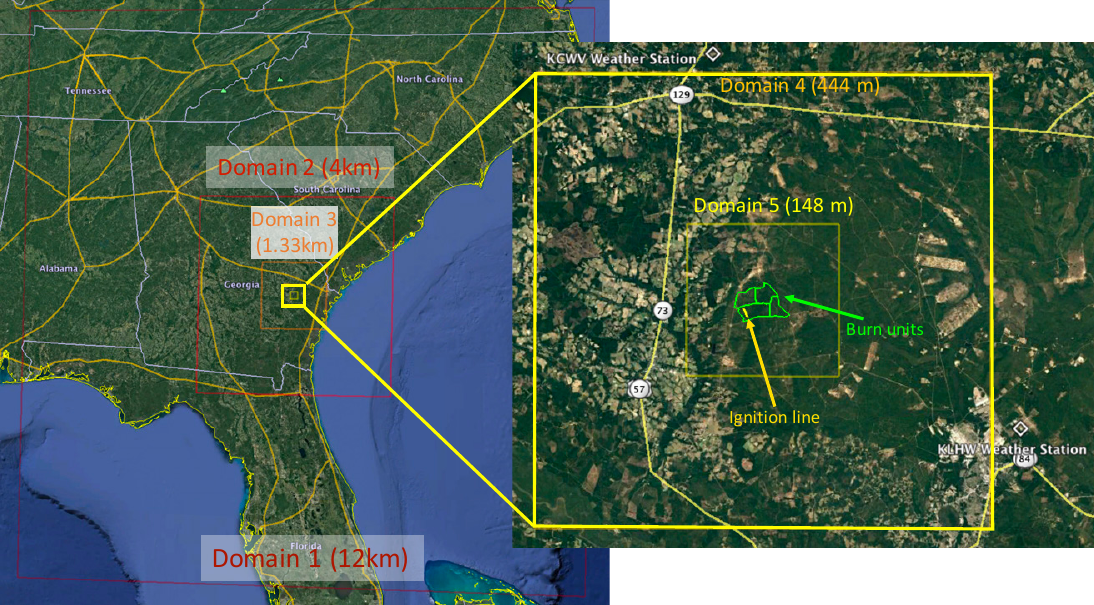}\\
(c)
\caption{WRF domain setups for burn simulation: (\textbf{a}) Fishlake, (\textbf{b}) North Kaibab, (\textbf{c}) Fort Stewart, with indication of nearby meteorological stations used to define typical days.}
\label{fig:3}
\end{figure}

Numerical simulations for the planned experimental burns were performed for the typical days (meeting burn requirements for the specific sites) obtained from the analysis described in Section~\ref{sec:typical}, in particular, Figure~\ref{fig:7}. The numerical experiments were conducted for Fishlake, North Kaibab, and Fort Stewart burn sites, and all used a multiscale setup of telescopic domains of grid scales gradually decreasing from 12~km to 148~m. Each simulation was performed with 30~m-resolution topography, land use and fuel maps from LANDFIRE, and driven by time-varying large-scale weather forcing obtained from Northern American Regional Reanalysis \citep[NARR,][]{Mesinger-2006-NAR}. Model configurations used for these runs are presented in Table~\ref{tab:3}. The domain configurations are shown in Figure~\ref{fig:3}. 
The Fishlake burn has been started as specified in the burn plan by a line ignition following the local mountain crest (see Figure~\ref{fig:3} a). For North Kaibab burn, a point ignition was used, as the ignition procedure wasn't specified for this site at the time of our simulations. The Fort Stewart simulations were executed with point-, single-line, and multi-line ignitions in order to cover a range of possible ignition scenarios.     

We have used the WRFx system \cite{Vejmelka-2016-WRFx} to conveniently set up the simulations. WRFx can download the selected atmospheric product data for the initial and boundary conditions,
process the USGS topography data and LANFDFIRE fuel data for the simulation domain, set up, execute, and monitor the WRF-SFIRE simulation, and postprocess the output into geolocated images, and stream the results to a visualization server. 

WRF-SFIRE~\cite{Mandel-2009-DAW,Mandel-2011-WFM} evolved from CAWFE~\cite{Clark-1996-CAF,Clark-2004-DCA} and it has been a part of WRF since release 3.2 as WRF-Fire~\cite{Mandel-2011-WFM,Coen-2013-WCW}. After release 3.3, WRF-SFIRE has continued to develop outside of WRF release~\cite{Kochanski-2016-TIS,Mandel-2014-RAA}. The version in WRF release was recently selected as a foundation of the operational Colorado Fire Prediction System (CO-FPS)~\cite{Jimenez-2018-HRC} and the level set method was improved~\cite{Munoz-2018-AFS}.

\begin{table}
\caption{Parameters of the experimental burn simulations}
\label{tab:3}
\small % Font size can be changed to match table content. Recommend 10 pt.
\centering
\begin{tabular}{lccc}
\toprule\\
\textbf{Burn Site:}	&
\textbf{Fishlake} &
\textbf{North Kaibab}&
\textbf{Fort Stewart}\\
\midrule\\
Meteo forcing&	NARR&NARR&NARR\\
Number of domains&	5 (realistic)&5 (realistic)&5 (realistic)\\
Domain sizes(XYZ)&	$97\times97\times41$&$97\times97\times41$&$97\times97\times41$\\
Model top 	& 13.2km&13.2km&13.2km\\
\multirow{2}{*}{Horizontal resolution}	& 12km/4km/1.33km&12km/4km/1.33km&12km/4km/1.33km\\
&/444m/148m&/444m/148m&/444m/148m\\
Vertical resolution& 5.3m – 2233m&5.3m – 2233m&5.3m – 2233m\\
Fire mesh resolution&	29.6m&29.6m&29.6m\\
Total number of runs&	7&	4&	10\\
Ignition&	Helicopter & 	Point&	Straight Line/Point\\
\multirow{4}{*}{Simulation start} 
& 09.03.2014 00:00 UTC & 09.05.2008 00:00 UTC & 04.22.2014 00:00 UTC\\
& 09.11.2016 00:00 UTC & 09.19.2015 00:00 UTC & 04.27.2009 00:00 UTC\\
& 09.22.2012 00:00 UTC & 09.01.2001 00:00 UTC & 02.27.2013 00:00 UTC\\
& 09.26.2015 00:00 UTC & 09.02.2011 00:00 UTC & 02.26.2008 00:00 UTC\\
Ignition Time& 	15:00 UTC (9:00 local)&	15:00 UTC (9:00 local)&	15:00 UTC (11:00 local)\\
Simulation Length& 48h&48h&48h\\
Fire Output Interval &5min &5min &5min\\
Time step (d05)	& 0.5s	& 0.5s	& 0.74s\\
1hr fuel moisture	&6.0\%	&6.0\%	&15.0\%\\
10hr fuel moisture	&8.0\%	&8.0\%	&13.0\%\\
100hr fuel moisture	&9.0\%	&9.0\%	&13.0\%\\
1000hr fuel moisture	&12.0\%	&12.0\%	&18.0\%\\
\bottomrule
\end{tabular}
\end{table}

\subsection{Statistical Analysis of Typical Burn Dates}
\label{sec:typical}
Coupled weather-fire-chemistry simulations (Section~\ref{sec:simulations}) provide important technological support for burn planning, operation, and analysis. To be relevant, these simulations are initialized with weather conditions typical of the burn location. In this analysis, “typical” is defined by a statistical analysis procedure applied to weather-station data records.

To start, consider a weather state defined by a single quantity e.g., air temperature. Then, one might define a ``typical'' day as one that has temperature closest to the historical sample mean of the temperature from all days considered.  

\begin{figure}[tbp]
\centering
\includegraphics[width=14cm]{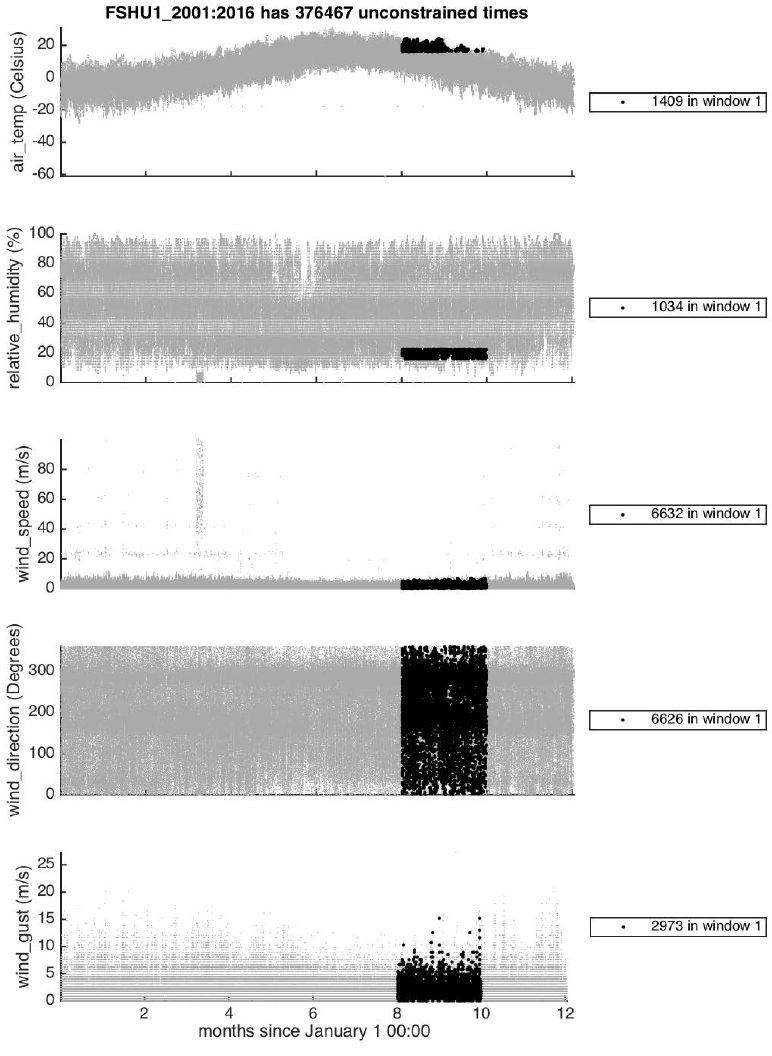}
\caption{Station data (vertical) vs month number (horizontal) for station FSHU1 (Fish Lake ranger station near Koosharem 10ENE, 38.55167\degree N, 111.72278\degree W, 8880 ft in Utah) in years 2001--2016. Points that do and do not meet the burn-requirements (Table~\ref{tab:req}) are dark and light gray, respectively, and enumerated in the legends and titles, respectively.
The abscissa is $t\pmod{12}$ for all the years indicated in the title.}
\label{fig:5}
\end{figure} 

More generally, a weather state is defined as a vector of temperature and multiple additional quantities, namely relative humidity, and the speed, direction, and gust of the horizontal wind vector.
Such data present a challenge, because the quantities have different units, value ranges, and statistical character, which make them incommensurate.
As an example, the input data (meteorological states comprising temperature, relative humidity, wind speed, direction and gust) for the ``typical day'' analysis are displayed in Figure~\ref{fig:5} for one of the stations.
Such figures are used as a ``sanity check'', i.e., a minimal quality-control of the data (indeed, some spurious values were discovered this way), as well as to visualize the burn requirements for state values, day-of-year and time-of-day, and the relative amounts of state values that do and do not meet the burn requirements. They also indicate that the variability of the raw data is far from that of a multivariate Gaussian random process.

\begin{table}[tbp]
\caption{Stations, time windows (mm/dd hh:00UTC) and value limits for potential burns (burn windows). $T$ – temperature, $\phi$ – relative humidity, $s$ – wind speed}
% Table 6
\label{tab:req}
\small % Font size can be changed to match table content. Recommend 10 pt.
\centering
\begin{tabular}{llcccccc}
\toprule
\textbf{Station} &
\textbf{Burn season} &
\textbf{min $T$} &
\textbf{max $T$} &
\textbf{min $\phi$} & 
\textbf{max $\phi$} & 
\textbf{min $s$} & 
\textbf{max $s$} \\
\midrule\\
FSHU1	& 09/01 16:00UTC - 10/31  18:00UTC& 61\degree F&85\degree F&16\% &22\% &	0& 15mph \\
KCWV&	02/20 16:00UTC -- 02/28  18:00UTC	&60	&90&	30&	55 & 6 & 20 \\
&	04/15 13:00UTC -- 04/30  14:00UTC	&60&	90	&30&	55	&6&	20\\
KLHW	&02/20 16:00UTC -- 02/28  18:00UTC	&60&	90&	30&	55&	6	&20\\
	&04/15 13:00UTC -- 04/30  14:00UTC	&60&	90&	30&	55&	6	&20\\
LCSS1&	 01/01 15:00UTC -- 02/01   19:00UTC	& 60&	90&	30&	55&	6	&20\\
	&12/01 15:00UTC -- 12/31 19:00UTC	&60	&90	&30	&55	&6	&20\\
QLBA3&	 09/01 17:00UTC -- 10/31 19:00UTC	&61	&85	&16	&22	&0	&15\\
TT084	& 09/01 16:00UTC -- 10/31 18:00UTC	&61	&85	&16	&22	&0	&15\\
\bottomrule
\end{tabular}
\end{table}

\begin{figure}[tb]
\centering
\includegraphics[width=12cm]{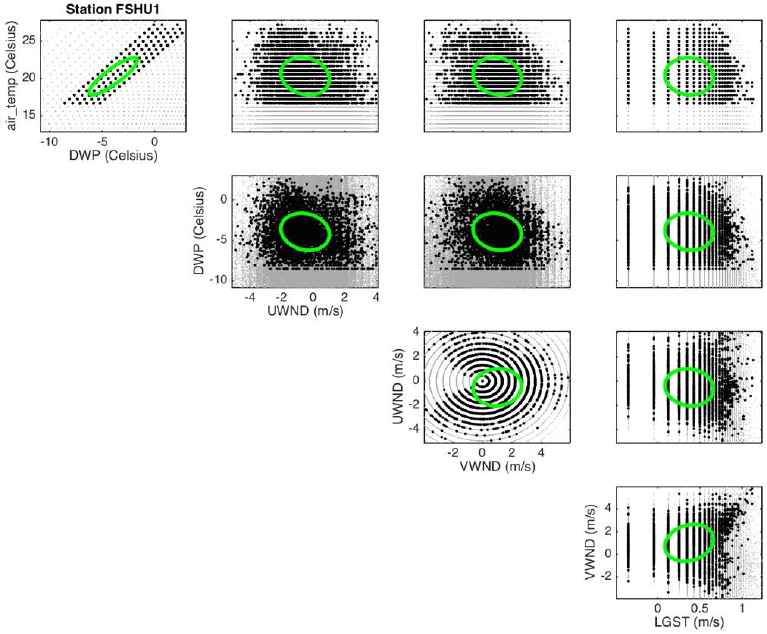}
\caption{Scatter plots of FSHU1 pairs of state variables. Black and gray indicate values that are and are not constrained by the burn requirements. The marginal covariance of a pair of variables is depicted by the green ellipse centered at the mean and with axes equal to the deviations in the principal directions. The ellipse projections on the coordinate axes have lengths equaling twice the corresponding marginal standard deviations. Ellipse area decreases with increasing correlation (if tilting up) or decreasing anti-correlation (if tilting down); a circle would indicate zero correlation. Only the values that satisfy the minimum and maximum values for burn requirements in Table~\ref{tab:req} are used to compute the covariances, in the considered years regardless of burn season.}
\label{fig:6}
\end{figure} 

For the analysis, the bounds on the value ranges are first removed by transforming from relative humidity, speed, direction and gust, to obtain the weather state column vector

\[x=\left[ T,{T}_{\textrm{d}},u,v,\ell  \right]^{\textrm{T}},\]
where $T$ is the temperature, ${T}_{\textrm{d}}$  is the dew point, $u$  and $v$  are eastward and westward components of the wind vector, and $\ell$  is the logarithm of the wind gust. The transformed data and their statistics are presented in Figure~\ref{fig:6}. It is evident from the scatter plots that the transformed state is still not very impressively Gaussian, but certainly is much more Gaussian than the original (temperature, relative humidity, wind speed, direction and gust) description. Also, the ellipse tilts and major-minor axis ratios confirm the general impression of the scatter of data points that meet the burn requirements.%

Even after this transformation, the weather-state vector entries still have different units, and their collection for all days considered varies more in some directions and less in others, so that measuring closeness by a simple root of sum of squared differences is not appropriate. Instead, we choose as typical days the days with the least Mahalanobis distance from the sample mean. The sample mean vector is computed as

\[\left\langle x \right\rangle =\frac{1}{N}\sum\limits_{t=1}^{N}{{{x}_{t}}},\]
where the sum is taken over all measurement times $t$ when $x_t$ satisfies the burn criteria (Table~\ref{tab:req}) in the considered years regardless of burn seasons. The sample covariance matrix is then computed by

\[C=\left\langle \left( x-\left\langle x \right\rangle  \right){{\left( x-\left\langle x \right\rangle  \right)}^{\text{T}}} \right\rangle \]
and the Mahalanobis distance of a vector \({{x}_{t}}\) from the sample mean is defined by

\[\left\| {{\delta }_{t}} \right\|=\left\| {{C}^{-1/2}}\left( {{x}_{t}}-\left\langle x \right\rangle  \right) \right\|=\sqrt{{{\left( {{x}_{t}}-\left\langle x \right\rangle  \right)}^{\text{T}}}{{C}^{-1}}\left( x_t-\left\langle x \right\rangle  \right)},\]
where the Cholesky factorization of $C$ and the Mahalanobis deviation vector at time $t$ are

\[{{C}^{1/2}}{{\left( {{C}^{1/2}} \right)}^{\text{T}}}=C, \qquad {{\delta }_{t}}={{C}^{-1/2}}\left( {{x}_{t}}-\left\langle x \right\rangle  \right).\]
Note that by definition, the %vectors
\({{\delta }_{t}}\) have zero sample mean and their sample covariance is identity, so the variance

\[\operatorname{var}\delta =\left\langle {{\delta }^{\text{T}}}\delta  \right\rangle = \left\langle \|\delta\|^2  \right\rangle =5,\]
which is useful to interpret the values of $\|\delta_t\|$.

The Mahalanobis distance puts more weight on the differences between weather state vectors along the directions where the collection of the weather state vectors varies less, on average. Another interpretation of this method is as a maximum likelihood: if the weather state vectors were a random sample from a multivariate Gaussian distribution, the days with smaller Mahalanobis distances from the sample mean would be the more likely ones.
Finally, this analysis does not directly consider steadiness i.e., what happens before or after the time $t$ that minimizes $\left\|\delta_t\right\|$. In theory, and to the extent the statistics are truly Gaussian, $\left\|\delta_t\right\|^2$ might be interpreted as a quadratic potential (like a harmonic oscillator has), so typical $x_t$ should be near equilibrium.

MATLAB software which implements the determination of typical days is provided in Supplementary Materials. 

\subsection{Statistical Optimization of Sensor Placement}
It is most useful to place (the limited number of) sensors where the observations or parameters of interest change or show the largest variance. To estimate the sensitivity of measurements to model parameters, a global sensitivity analysis with repeated Latin Hypercube Sampling \cite{McKay-1979-CTM,McKay-1995-EPU,Saltelli-2004-SAP} was used. Consider a simple precursor method: run a number of simulations with the values of the parameters of interest chosen randomly and then compute the sample variance of the output over the simulations at various potential sensor locations.  The method utilized here improves on such a conventional approach in two aspects: repeated Latin Hypercube Sampling (rLHS) allows for a drastic reduction in the number of simulations to be made by a systematic coverage of the parameter space, and the Sobol variance decomposition \cite{Sobol-2001-GSI} can distinguish the effect of individual parameters. This method was originally developed to evaluate computer simulations at the Los Alamos National Laboratory \cite{McKay-1979-CTM,McKay-1995-EPU} and was recently used for guiding sensor placement for oil reservoir engineering \cite{Chugunov-2014-TSD}.

\subsubsection{Repeated Latin Hypercube Sampling and Sobol Variance Decomposition}
\label{sec:rLHS}

Consider a model with \(L\) parameters, divide the range of each parameter into \(N\) intervals of equal probability, and for each interval choose the probability midpoint as a sampling point \(1,\ldots ,N\). Then set up \(N\) parameter vectors by choosing the sampling points of each of the \(L\) parameter randomly, without repeating the same sampling point in any of the parameters. The simulation is run on each parameter vector. See Figure~\ref{fig:LHS} for an example.

\begin{figure}[tb]
\small % Font size can be changed to match table content. Recommend 10 pt.
\centering
\fbox{
\begin{tabular}{ccccc}
%\toprule\\
%\multicolumn{5}{c}{\textbf{Sampling points}}\\
%\midrule\\
3&	1&  2&	4&	5\\
3&	1&	5&	4&	2\\
4&	3&	2&	1&	5\\
3&	5&	2&	1&	4\\
1&	3&	2&	4&	5\\
4&	5&	1&	3&	2\\
5&	2&	4&	3&	1\\
%\bottomrule
\end{tabular}
}
\caption{Latin Hypercube Sampling (LHS). Each of $L=7$ parameters takes the values from its $N=5$ sampling points. The rows are random permutations of the numbers $1$ to $5$, while the columns are the parameter vectors on which the simulations will be run.}
\label{fig:LHS}
\end{figure}

Repeated Latin Hypercube Sampling (rLHS) consists of \(r\) repeats of LHS, with new random permutations of the choices of the sampling points for each parameter. The complete rLHS scheme requires \(rN\) simulations, with each simulation receiving a vector of \(L\) parameters. 

Each of the \(rN\) simulations delivers its output \(Y\). Now fix the value of parameter \({{X}_{i}}\) at one of the sampling points \({{x}_{ij}}\) and take the average of the outputs \(Y\) over all simulations where parameter had that value, \({{X}_{i}}={{x}_{ij}}\). The result is a function of the value of the parameter \({{X}_{i}}\). This function is called the conditional expectation of \(Y\) given \({{X}_{i}}\), and we denote it by \({{Y}_{i}}\). Since the parameter \({{X}_{i}}\) attains \(N\) sampling values with equal probability \(1/N\), the conditional mean \({{Y}_{i}}\) also attains \(N\) possible values with equal probability \(1/N\) each. From those values of \({{Y}_{i}}\), we compute its sample variance \(\operatorname{var}\left( {{Y}_{i}} \right)\) in the usual way except that we use the denominator \(N\) instead of \(N+1\) in the expression for the variance. We also compute the sample variance \(\operatorname{var}\left( Y \right)\) over all \(rN\) values, again using \(rN\) in the denominator instead of the usual \(rN+1\).

The importance of this procedure lies in the following interpretation:
\begin{itemize}[leftmargin=*,labelsep=4mm]
\item	$\operatorname{var}\left( Y \right)$ is the total variability of the output $Y$ over the range of the parameters ${X}_{1},\ldots,{X}_{L}$. We take this as the strength of the signal from the sensor.
\item	$\operatorname{var}\left( {{Y}_{i}} \right)$, called the Variance of the Conditional Expectation (VCE), is an estimate of the variability of the output $Y$ due to the parameter $X_i$.
\item The ratio 

\[\operatorname{eff}\left( {{Y}_{i}} \right)=\operatorname{var}\left( {{Y}_{i}} \right)/
\operatorname{var}\left( Y \right)\] 
is an estimate of the relative effect of parameter ${X}_{i}$ on the output $Y$. It is called the correlation ratio \cite{McKay-1995-EPU} or the sensitivity index \cite{Saltelli-2004-SAP} of the parameter ${X}_{i}$.
\end{itemize}

This interpretation is motivated by the case when the model is linear,

\[Y={{a}_{1}}{{X}_{1}}+\cdots +{{a}_{L}}{{X}_{L}},\]
where the coefficients ${a}_{1},\ldots ,{a}_{L}$ are constant and the inputs ${{X}_{1}},\ldots ,{{X}_{L}}$ are independent random variables with unit variance. Then, it can be shown that under suitable assumptions (which are certainly satisfied e.g., when ${{X}_{i}}$ have Gaussian distribution), the VCE from the rLHS procedure with Sobol’s variance decomposition provides asymptotically exact approximation of the VCE of $Y$ given ${{X}_{i}}$ for large number of samples and intervals: 

\begin{equation}
\operatorname{var}\left( {{Y}_{i}} \right) %=\operatorname{var}\operatorname{E}\left( Y|{{X}_{i}} \right)
\to a^2_i  \label{eq:var}
\end{equation}
as  the number of repetitions $r\to \infty $ and the number of intervals $N\to \infty $.
That is, \emph{the method recovers asymptotically the variance of the output $Y$ due to one parameter $X_i$ from the signal $Y$ that combines the effect of many independent parameters, by probing the model.} If the parameters are not independent, or the model is not linear, the recovery is only approximate. 

MATLAB software which implements the rLHS and validates the implementation by observing the convergence in (\ref{eq:var}) computationally is provided in Supplementary Materials. 

In the present application, $Y$ is defined as one of measured values at some location e.g., the vertical wind velocity at a fixed time, longitude, latitude, and height. If the effect of one particular parameter is of interest, it makes sense to look for locations where the sensitivity index of that parameter is large. Then the signal due to that parameter stands out from the rest.

In the case where there is a small number of repetitions and intervals, the convergence of the sample variance of the measured value $Y$ as well as the convergence in (\ref{eq:var}) may be far from the asymptotic range. But, since the measured values $Y$  are taken from the same set of simulations and, in particular, when the measured values $Y$ are the same measured quantity at nearby locations, their sampling errors are correlated and comparing their sample variances as in the figures in Section~\ref{sec:simulations-results} is justified.

Finally, since the quantities of interest will be associated with a fixed location while the fire is moving, there is only a significant signal when the fire or smoke is advected over the sensor location. Hence,  the sums of time series of variances associated with a fixed location, which represents the total variability captured by the sensor over time are also of an interest.

\section{Results}
\subsection{Determination of Typical Burn Days}

\begin{figure}[tbph]
\centering
\includegraphics[width=14cm]{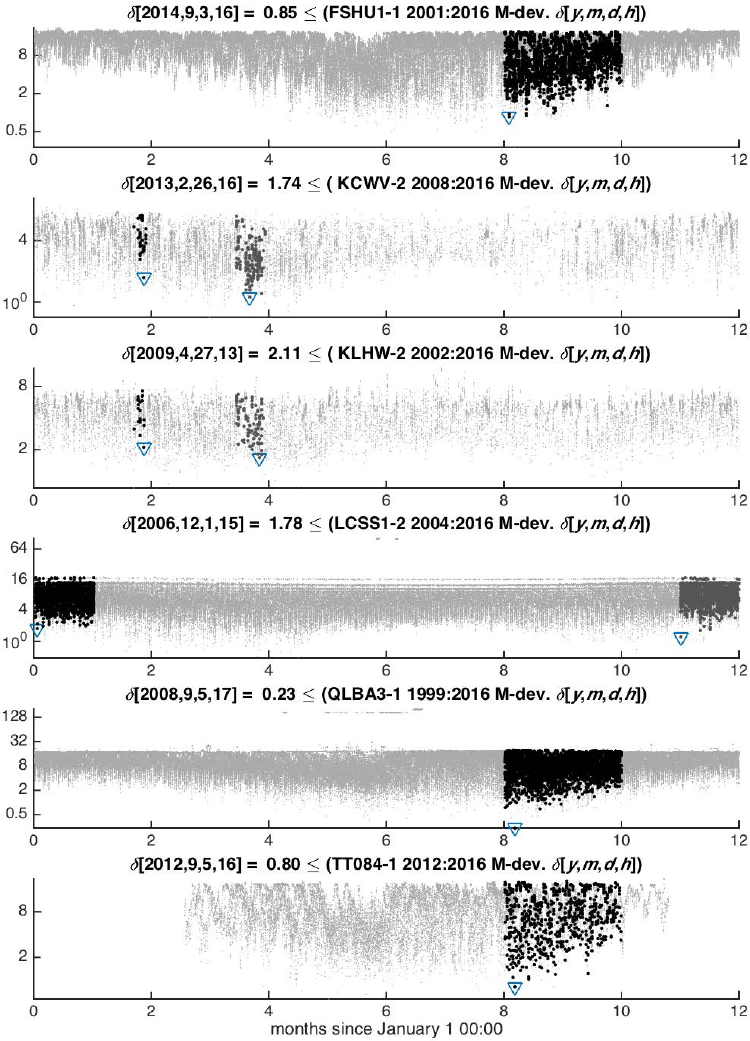}
\caption{Time series of Mahalanobis (standardized) deviation norm $\|\delta_t\|$ for each station (6 rows). Points for states that do and do not meet burn requirements are dark and light gray, respectively. For each station, the most typical day in the burn window has the smallest $\|\delta_t\|$ and is listed in the title and indicated with a $\triangledown$ marker. The abscissa is $t\pmod{12}$ for all the years indicated in the plot titles.}
\label{fig:7}
\end{figure}

From Section~\ref{sec:typical}, the Mahalanobis distance \(\left\| {{\delta }_{t}} \right\|\) of the weather vector at a station from the long-term mean is dimensionless, its mean value squared is \(\left\langle\left\| {\delta } \right\|^2\right\rangle =5\), and the most ``typical'' day is defined as one with the smallest \(\left\| {{\delta }_{t}} \right\|\) in the given burn time window. The resulting identification of typical burn days is in Figure~\ref{fig:7} and Table~\ref{tab:typical}.

\begin{table}[tbph]
\caption{Summary of most typical days identified for the burn sites.}
\label{tab:typical}
\small % Font size can be changed to match table content. Recommend 10 pt.
\centering
\begin{tabular}{cccccc}
\toprule
\textbf{Burn site} & \multicolumn{2}{c}{\textbf{Fishlake}} &
\textbf{North Kaibab} & 	\multicolumn{2}{c}{\textbf{Fort Stewart}} \\
\textbf{Weather station} &	
\textbf{FSHU1} &
\textbf{TT084} &
\textbf{QLBA3} &
\textbf{KCWV} &
\textbf{KLHW} \\
\midrule
\textbf{Typical} &	2014-09-03&	2012-09-05&	2008-09-05&	2014-04-22&	2009-04-27\\
\textbf{days} &	2015-09-26&	2012-09-04&	2005-09-19&	2013-02-24&	2013-02-26\\
\bottomrule
\end{tabular}
\end{table}

\subsection{Results from Numerical Experiments}
\label{sec:simulations-results}

\subsubsection{Simulations of planned experimental burns in Fishlake}
\label{sec:Fishlake}

The analysis of the expected plume characteristics has been performed for the Fishlake simulations. Three simulations were run for the day of 09.03.2014, (which has been identified based on the weather observations from the FSHU1 weather station as the most typical day meeting the burn criteria defined in Table~\ref{tab:req}): 
\begin{enumerate}
\item reference simulation without fire,
\item standard simulation with fire, and
\item simulation with fire and doubled heat flux from the fire to the atmosphere. 
\end{enumerate}
The first run serves as a reference point for assessing the impact of the fire on the boundary layer when compared to the second run. The third run is intended to account for possible underestimation of the fire heat fluxes under high fuel load conditions characteristic for the Fishlake burn site, as well as to assess sensitivity of the simulated plume height and vertical velocities to the fire heat flux. 

\begin{figure}[tb]
\centering
\begin{tabular}{cc}
\includegraphics[width=7cm]{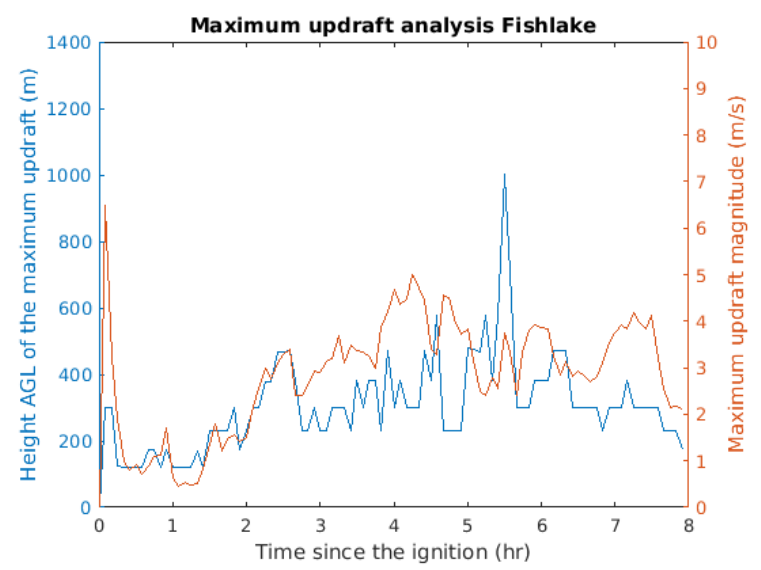}  &
\includegraphics[width=7cm]{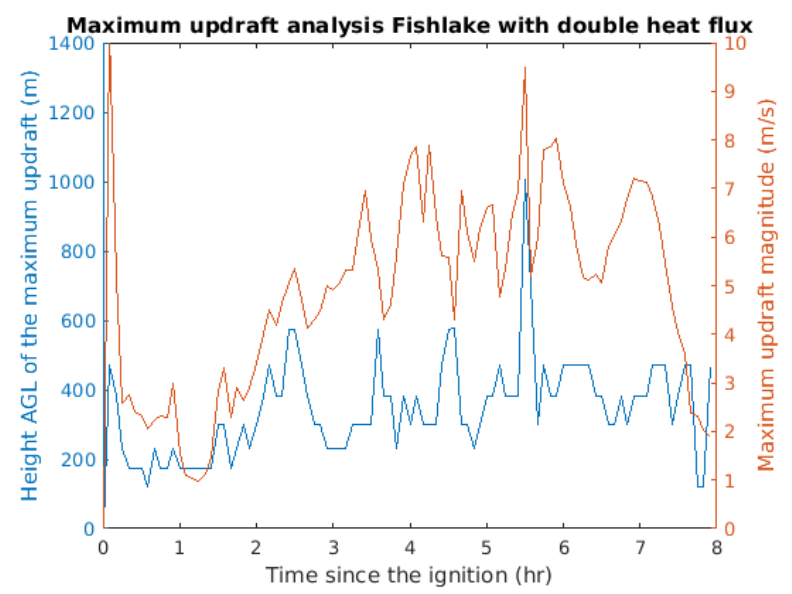} \\
(a) & (b) \\
\end{tabular} 
\caption{Time series of the maximum updraft strength and location from the Fishlake simulation for 09.03.2014. (\textbf{a}) Standard simulation, (\textbf{b}) Simulation with doubled fire heat flux released to the atmosphere.}
\label{fig:11}
\end{figure} 

\begin{figure}[tb]
\centering
\begin{tabular}{cc}
\includegraphics[width=7cm]{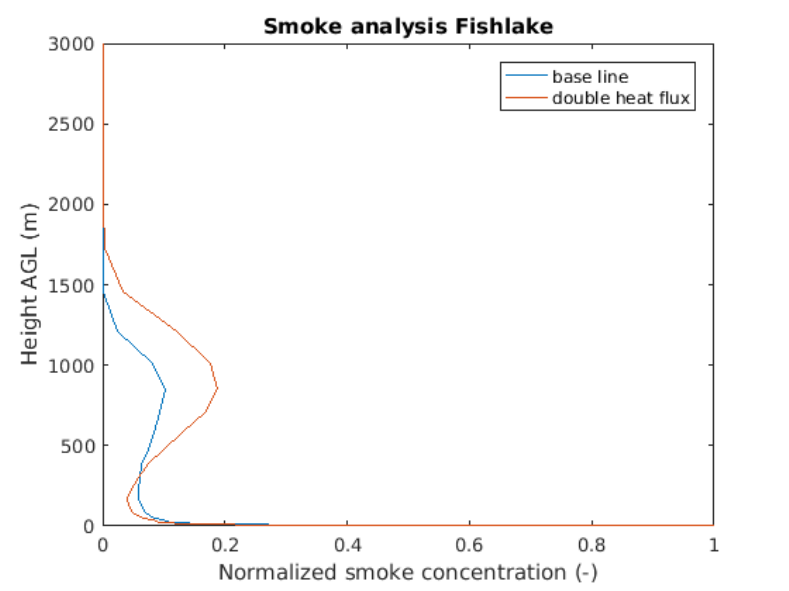}  &
\includegraphics[width=7cm]{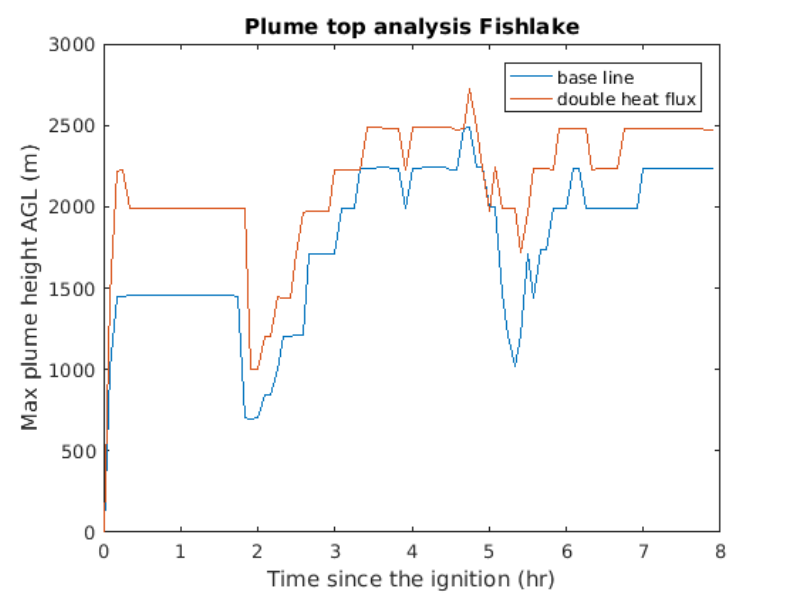} \\
(a) & (b) \\
\end{tabular} 
\caption{Smoke analysis for the Fishlake burn executed for 09.03.2014. (\textbf{a}) Vertical profile of normalized smoke concentration. (\textbf{b}) Time series of the plume top height.}
\label{fig:12}
\end{figure} 

As shown in Figure~\ref{fig:11}, for the standard and doubled heat flux fire simulations, maximum updraft velocities tended to spike during the ignition and then gradually rise over the next 4 hours.  The maximum updraft velocities reach 5 m/s in the standard simulation and twice as much in the simulation with the doubled heat flux. The elevations where maximum updrafts occur are significantly less sensitive to the fire heat flux enhancement. In both cases, they vary between 170m and 1000m. These simulations suggest that from the standpoint of the plume characterization it is important that the experimental burn lasts long enough so that the smoke column fully evolves (at least a couple of hours). The vertical in-plume velocity measurements (LIDAR vertical scans, UAV platforms) should allow for plume scanning up to at least 1.5 km and be able to withstand vertical velocities up to 10m/s. In order to provide a guideline for the in-plume measurements of chemical species, the simulations were executed with fire smoke represented as a passive tracer. Figure~\ref{fig:12}(a) shows vertical profiles of the smoke concentration normalized by its maximum value for the baseline simulation and the case with doubled heat flux. Both cases show similar concentration profiles with a maximum near the ground and a single peak located around 800m above the ground level. In the double-heat case smoke penetrates into higher elevations. The time series in Figure~\ref{fig:12}(b) show a rapid plume rise during the ignition, followed by a 2h period of constant plume top height, followed by a sudden drop and a gradual increase with some further fluctuations. As the heat released during the ignition procedure is also doubled in the double heat flux case, the differences in the plume top heights between the cases are most evident during the first 2h after ignition. Later, as the plume fully develops, the plume heights in these two simulations become similar (within 250m). The concentration plots presented in Figure~\ref{fig:12}(a) suggest that plume sampling should be focused on the layer 3000m AGL (Above Ground Level), with a particular focus on the 500m and 1500m AGL, where the simulations tend to produce the highest tracer concentrations. It has to be noted that as the plume evolves in response to the changing  atmospheric conditions and the fire behavior the elevation of the plume layer may change rapidly.

\begin{figure}[tb]
\centering
\begin{tabular}{cc}
\includegraphics[width=7cm]{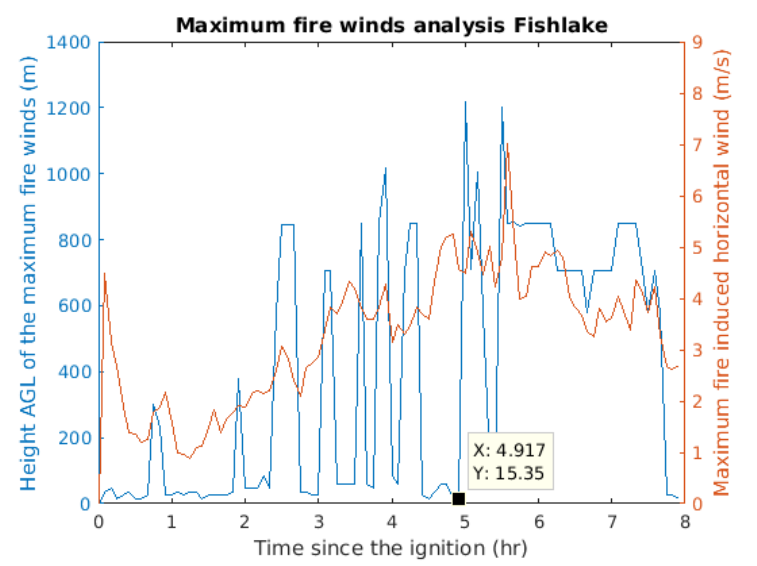}  &
\includegraphics[width=7cm]{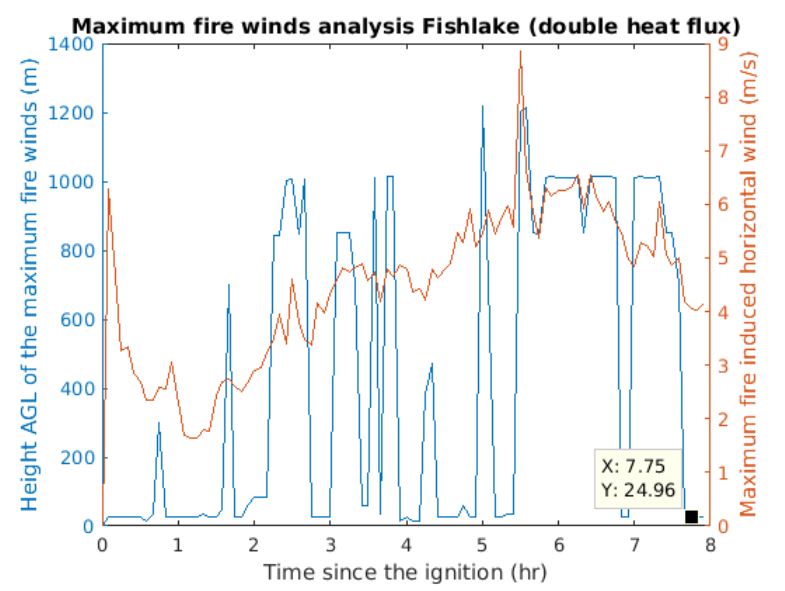} \\
(a) & (b) \\
\end{tabular} 
\caption{Analysis of the fire-induced winds for the Fishlake simulation executed for 09.03.2014. (\textbf{a}) Time series of the maximum fire-induced horizontal winds and their heights in the baseline case.  (\textbf{b}) Same in the double heat case.}
\label{fig:13}
\end{figure} 

In order to assess the fire's impact on boundary layer flow in the baseline and double heat cases, the horizontal fire-induced winds are computed as a difference between the U and V wind speed components from the fire runs and the reference no-fire run. The time series of the levels where the maximum fire-induced winds occur and their magnitudes for both fire simulations are presented in Figure~\ref{fig:13}. The elevations of the fire-induced wind maxima are similar in both simulations, suggesting that the level at which the fire accelerates flow most rapidly is not particularly sensitive to the fire intensity (heat flux). However, the magnitudes of the fire-induced wind are. The maximum winds in the baseline simulation reach 7 m/s, while in the double heat flux case they reach 9m/s. The simulations indicate that the maximum wind flow is either close to the ground (15-25m AGL) or aloft, mostly between 800m and 1200m AGL. Therefore, from the measurement standpoint, the optimal sampling strategy would contain a combination of a meterological tower sampling the near-surface flow and Doppler lidar(s) providing wind speed measurements at elevations at least up to 1500m AGL.

\subsubsection{Simulations of planned experimental burns in North Kaibab}
\label{sec:NorthKaibab}

\begin{figure}[tb]
\centering
\begin{tabular}{cc}
\includegraphics[width=7cm]{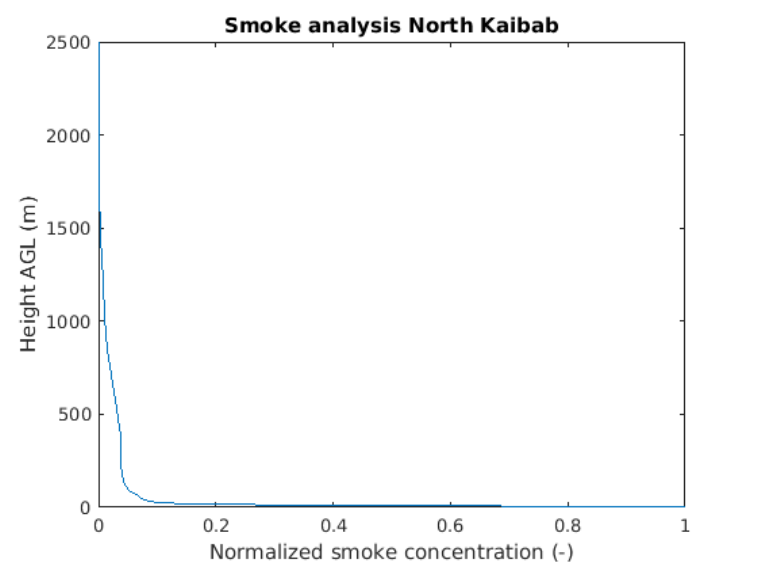}  &
\includegraphics[width=7cm]{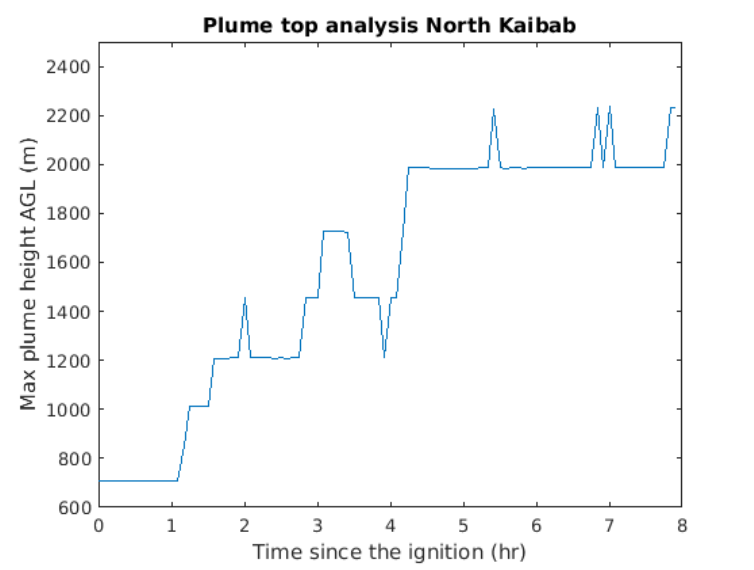} \\
(a) & (b) \\
\end{tabular} 
\caption{Smoke analysis for the North Kaibab burn simulation executed for 09.05.2008. (\textbf{a}) Vertical profile of normalized smoke concentrations.  (\textbf{b}) Time series of the simulated plume top height.}
\label{fig:14}
\end{figure} 

To provide insight into the second western burn at the North Kaibab site, a simulation was performed for day 09.05.2008, identified based on the weather observations from the QLBA3 weather station as the most typical day meeting the burn criteria defined in Table ~\ref{tab:req}. In the North Kaibab case, Figure~\ref{fig:14}(a) shows that the smoke layer is more uniformly distributed across the first 1500m than in the Fishlake case, with no evident concentration peak. 
Figure~\ref{fig:14}(b) shows lower simulated plume top heights than in the Fishlake case. They increase gradually over the first 4 hours and stabilize at around 2000m, 200--700m lower than in the Fishlake runs. 

\begin{figure}[tb]
\centering
\begin{tabular}{cc}
\includegraphics[width=7cm]{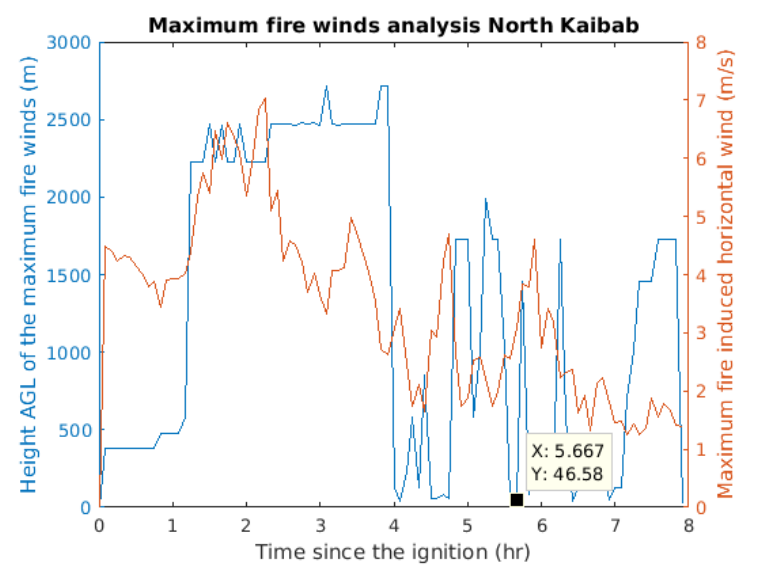}  &
\includegraphics[width=7cm]{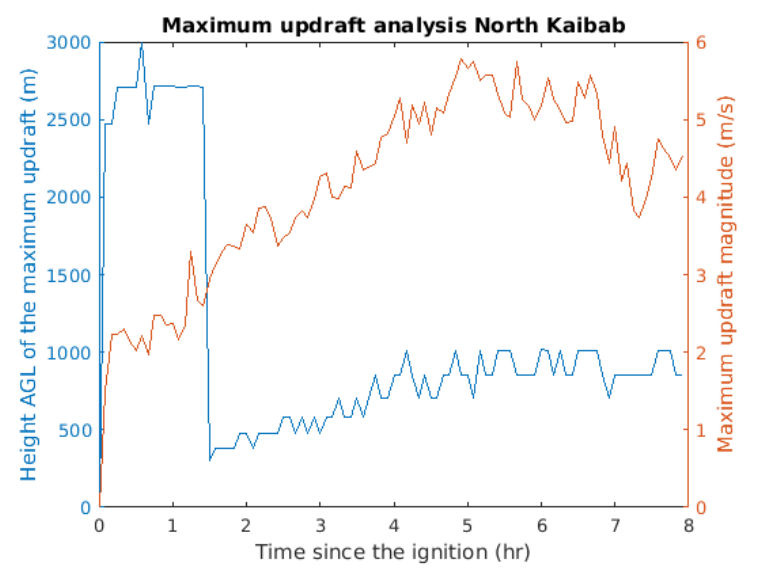} \\
(a) & (b) \\
\end{tabular} 
\caption{Analysis of the fire-induced wind and updraft for the North Kaibab simulation executed for 09.05.2008. (\textbf{a}) Time series of the maximum fire-induced horizontal winds.  (\textbf{b}) TTime series of the maximum fire-induced updraft.}
\label{fig:15}
\end{figure} 

The evolution of the fire-induced flow is presented in Figure~\ref{fig:15}. The fire-induced winds, in this case, are similar in magnitude to the Fishlake baseline case, but their time evolution is different. Contrary to the winds from the Fishlake simulation showing a strong initial spike in fire-induced winds followed by a gradual increase, the North Kaibab winds ramp up to about 4m/s, reaching a maximum of about 7 m/s at 2.5h into the simulation, and then keep decreasing. The timing of the peak surface winds corresponds to the time when the maximum updraft is located relatively close to the surface (below 500m AGL). The convective updraft evolves gradually in time, reaching maximum values close to 6m/s after 5h since ignition. Note that, as a result of a point ignition (less intense than the line ignition in the Fishlake burn), this evolution is not disturbed by the initial ignition-induced spike evident in the Fishlake run (Figure~\ref{fig:13}). The updraft analysis shown in  Figure~\ref{fig:15}(b) indicates that over the first one-and-a-half hours, the maximum updrafts are located at 2500-3000m AGL (probably as an effect of fire-induced disturbance inducing vertical mixing at high elevations). As the fire evolves, updrafts in the plume core intensify and become dominant, which is indicated by the elevations of the maximum updrafts going back to 500-1000m AGL.

 \subsubsection{Simulations of planned experimental burns in Fort Stewart}
\label{sec:FortStewart}

\begin{figure}[tb]
\centering
\includegraphics[width=15cm]{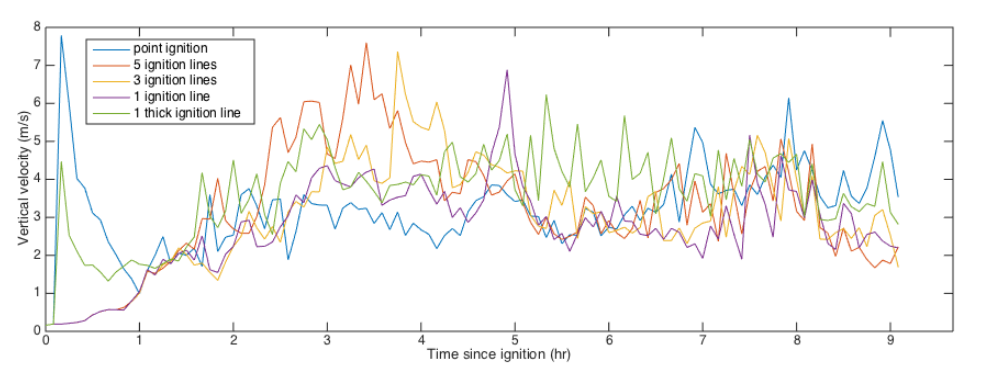}  \\
(a) \\ \ \\
\includegraphics[width=15cm]{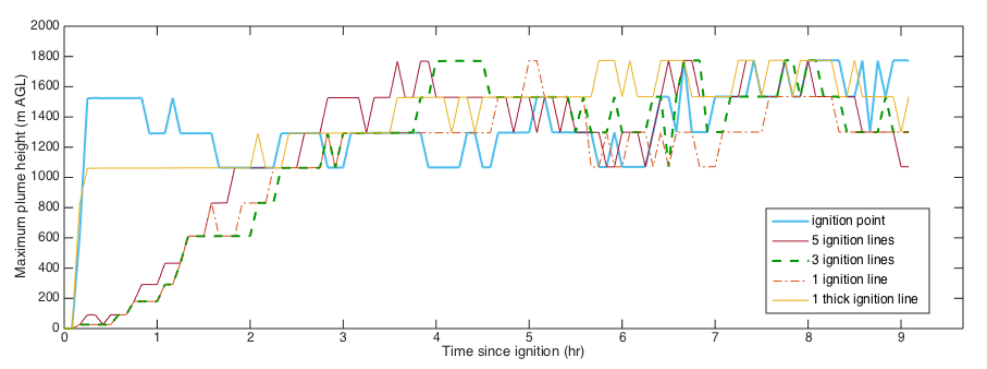} \\
 (b) \\
\caption{Results from numerical simulations for Fort Stewart (2013.02.14) with different ignition procedures. (\textbf{a})  Time series of the simulated maximum fire-induced updraft. (\textbf{b}) Same for plume top height.}
\label{fig:16}
\end{figure} 

Numerical simulations for Fort Stewart were performed with the main intent of providing an insight into the impact of the ignition procedure on plume evolution. For prescribed burns that require complex aerial ignitions, multiple ignition lines are formed in a relatively short period of time, with significant heat fluxes coming directly from the igniting agents like the gasoline mixture in Helitorch or ping-pong balls (DAID) systems. As the aerial ignition procedure is generally fast and difficult to precisely capture by scanning IR systems which do not provide a continuous static field of view, it is important to know how important the ignition itself is for further plume evolution. In order to assess that, 5 different ignition procedures were simulated: a single ignition point, a single ignition line of two different thicknesses, a set of 3 parallel lines, and a set of 5 parallel lines. All ignition lines were oriented approximately from northwest to southeast (perpendicular to the mean wind direction). They were 2 m thick, except for the single bold ignition line which was 30m thick. The lateral rate of spread within the ignition lines was set to 0.5 m/s in order to assure gradual initial heat release. Each line was ignited to its full length in 200 s. In the multiple-line cases, ignitions were performed sequentially, and the spacing between the lines was 250m. The results from these simulations are presented in Figure~\ref{fig:16}. The time series of the maximum vertical velocity and the plume height indicate that the ignition itself plays a significant role in updraft evolution, especially during the initial phase of the burn. The time series from different ignition procedures only start to converge 7 hours after ignition. These results suggest that for relatively short experimental fires (expected to last a couple of hours), accurate documentation of the ignition procedure is critical for a realistic representation of the initial plume evolution in subsequent numerical simulations.  

It should be also noted that the initial interactions between ignition lines are hard to account for in numerical experiments performed using models with parameterized fire spread. In these models the ignition process itself is crudely simplified, limiting their capabilities in terms of rendering these effects. Therefore, simple ignition patterns are desired in experimental burns so that the fire can be realistically initialized in the model.

\begin{figure}[tb]
\centering
\begin{tabular}{cc}
\includegraphics[width=7cm]{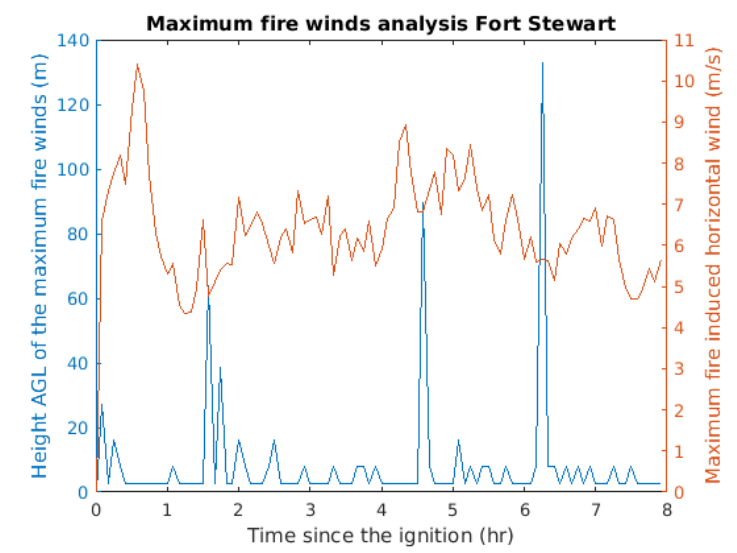}  &
\includegraphics[width=7cm]{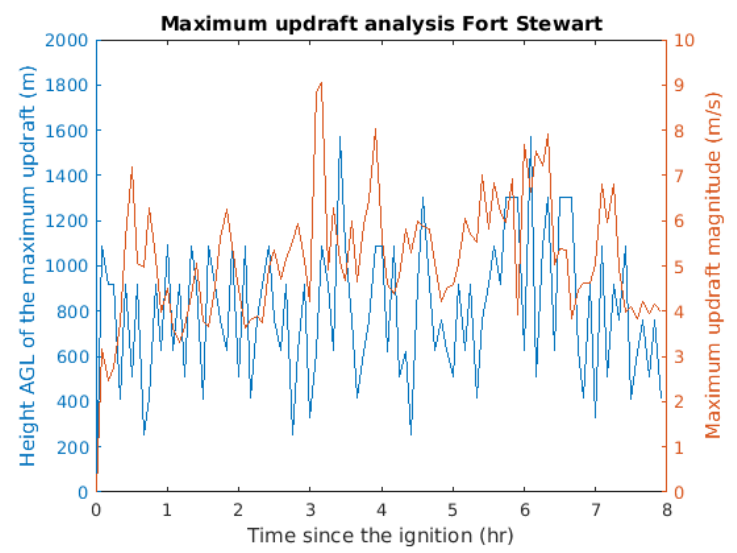} \\
(a) & (b) \\
\end{tabular} 
\caption{Analysis of the fire-induced wind and updraft for the Fort Stewart simulation executed for 04.22.2014. (\textbf{a}) Time series of the fire-induced maximum horizontal wind speed.  (\textbf{b}) Same for the updraft.}
\label{fig:17}
\end{figure} 

Aside from the simulations targeting the ignition effects, additional simulations were performed for 04.22.2014 (a typical day based on statistical analysis of observations from the KCWV weather station (Figure~\ref{fig:3}(c)). Results from this numerical experiment are presented in Figure~\ref{fig:17}. The Fort Stewart site is significantly different from south-western Fishlake and North Kaibab sites in terms of both fuel and topography. The latter ones are located in areas of complex terrain where steep elevation strongly impacts both vertical and horizontal flows, making fire-induced flows harder to detect. It seems that in complex terrain the destabilization induced by the fire can impact the atmospheric state at significant distances downwind from the plume core. Strong upslope winds can induce updrafts of a similar magnitude to the fire itself, and the interaction between the flow and the mountain ridges may lead to significant variations in the horizontal winds that become visible only when differences between fire and no-fire cases are computed. As a consequence, the heights of maximum fire-induced winds and vertical velocities fluctuate strongly in the North Kaibab simulations but are relatively steady in the Fort Stewart burn. In Fort Stewart for instance, the elevations of strongest fire-induced winds are confined to a shallow layer between 0 and 140m AGL, with the majority of strongest fire winds occurring within the 20m layer above the ground (Figure~\ref{fig:17}(a)). These results suggest that a combination of meteorological towers and LIDARs located close to fire line should be able to sample most of the maxima in the fire-induced flow. In the Fort Stewart simulation, both fire-induced winds and updraft velocities evolve very quickly, reaching a quasi-steady state just after one hour from the ignition. These simulations suggest, that despite the fuel load and expected fire intensity are lower for Fort Stewart than for Fishlake burns, in the context of characterization of the fire-induced circulation, the Fort Stewart site has a potential to provide better and easier to interpret results relative to the more complex southwestern sites.

% \afterpage{\clearpage}
 
\subsection{Sensitivity Study of Sensor Placement}
\label{sec:sensitivity}

We have performed a sensitivity study following the rLHS with variance decomposition methodology, described in 
Section~\ref{sec:rLHS}. We have chosen the Fishlake simulation standard case from Section~\ref{sec:Fishlake} as the base case.

\subsubsection{Sampling Setup}
\label{sec:sampling}
We have chosen $L=7$ simulation parameters, which were:
\begin{enumerate}
\item 10h-fuel moisture content, varying from 0.04 to 0.14 water mass/dry fuel mass. 1h-fuel fuel moisture was 10h-fuel moisture minus 0.01, 100h-fuel moisture was 10h-fuel moisture plus 0.01f, and live fuel moisture was 0.78. These values were entered as initial moisture values and did not change with time. 
See \cite{Mandel-2012-APD,Mandel-2014-RAA} for a further description of the fuel moisture in WRF-SFIRE.
\item Heat extinction depth, varying from 6m to 50m. The fire heat flux is entered into WRF boundary layer with exponential decay, rather than all into the bottom layer of cells. The heat is apportioned depending on the height of the cell center above the terrain, with weight 1 at the ground and   at the heat extinction depth. This gradual heat insertion is a parameterization for unresolved mixing and radiative heat transfer.
\item	Heat flux multiplier, varying from 0.5 to 2. The heat flux multiplier was chosen as a measure of the fire effect on the atmosphere; unlike the fuel load, it does not influence the rate of spread.
\item Multiplier for the omnidirectional component $R$ in the approximate fire rate of spread (ROS) formula following \cite{Rothermel-1972-MMP},

\[ROS = R + R_W + R_S\]
\item Multiplier for the wind-induced ROS component  $R_W$
\item Multiplier for the slope-induced ROS  component $R_S$
\item The simulation day, selected from the "typical" burn days 
\end{enumerate}
The first 6 parameters were constant over the whole fire simulation domain. Since the 6 parameters are positive, first they were transformed by taking the logarithm and then a Gaussian distribution centered within the transformed interval was determined so that 90\% probability was between the lowest and the highest value. The transformed interval was then divided into subintervals of equal probability, sampling points chosen as the middle probability values, and transformed back into the original physical space. The resulting sampling values are in Table~\ref{tab:sampling}. 
 
\begin{table}[tbph]
\caption{Sampling points for Latin Hypercube Sampling.}
\label{tab:sampling}
\small % Font size can be changed to match table content. Recommend 10 pt.
\centering
\begin{tabular}{lccccc}
\toprule
\textbf{Sampling point} &	
\textbf{1} &
\textbf{2} &
\textbf{3} &
\textbf{4} &
\textbf{5} \\
\midrule
10-fuel moisture (kg/kg)&	0.0459	&0.0613	&0.0748	&0.0914&	0.1219\\
Heat extinction depth (m)&	7.5830&	12.3531&	17.3205	&24.2854&	39.5620\\
Heat flux multiplier (1)&	0.5827&	0.8017&	1.0000&	1.2473&	1.7161\\
Multiplier for $R$ (1)	 & 0.5827	&0.8017	&1.0000	&1.2473	&1.7161 \\
Multiplier for $R_W$ (1)	& 0.5827	&0.8017	&1.0000	&1.2473	&1.7161\\
Multiplier for $R_S$  (1)	& 0.5827	&0.8017	&1.0000	&1.2473	&1.7161\\
Simulation day (date)	&
09.03.2014	&
09.11.2016	&
09.22.2012	&
09.26.2015	&
09.27.2015\\
\bottomrule
\end{tabular}
\end{table}

Total $r=200$ repetition was done, resulting in $rN=1000$ simulations. The selection of the sample points in the first replicant is in Figure~\ref{fig:LHS}.
For the purpose of the sensitivity analysis, the output of the model is the value of a quantity of interest at a location in the physical space (a node of the simulation mesh on the Earth's surface, or interpolated to a specified height). We then visualize the values associated with the quantity of interest, such as the VCE (Variance of the Conditional Expectation, an estimate of the variability of the output due to the parameter change, cf., Section~\ref{sec:rLHS}) as a function of location on the Earth surface. Note that in WRF, the vertical position of mesh nodes is not fixed, rather it is derived from the geopotential height, which is a part of the solution, and it changes with time.

We consider the following quantities of interest:
\begin{enumerate}
\item	The vertical velocity vector component $W$, interpolated to a given height above the terrain, or to a given altitude above the sea level.
\item	The smoke intensity (the concentration of WRF tracer \texttt{tr17\_1}), interpolated to a given height above the terrain, or to a given altitude above the sea level.
\item	Plume-top height, derived from the smoke concentration.
\end{enumerate}

Because some extreme values of the sampling parameters caused instability in the WRF surface scheme during and immediately after ignition (when the updraft carrying the heat away is not yet established), we have simulated the initial 30 minutes when the ignition occurs using the standard case and only then switched to the modified parameter values. We show the results from simulations running for an additional 7.5 hours, during which the model state was captured and analyzed at hourly intervals. The resultant 7 time frames were averaged to provide a picture corresponding to the 7h of the burn.

\subsection{Computational results for the Fishlake burn simulation}

\begin{figure}[tb]
\centering
\begin{tabular}{cc}
\multicolumn{2}{c}{\includegraphics[width=15cm]{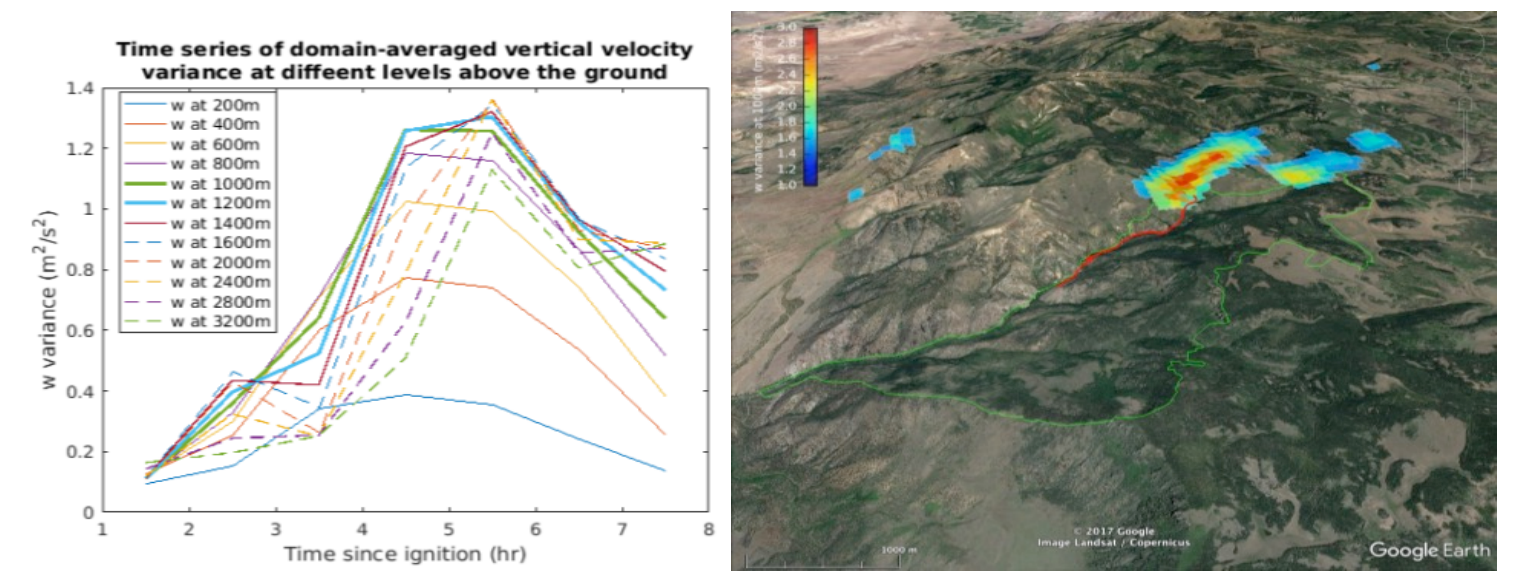}}\\
\makebox[7cm][c]{(a)} & \makebox[7cm][c]{(b)} \\
\end{tabular}
\caption{(\textbf{a}) The time series of the variance of vertical velocity at different heights above the ground from simulations executed with parameters from the rLHS sampling. (\textbf{b}) Map of the variance of the vertical velocity at 1200m with the indication of the burn plot (green contour) and the ignition (red line.}
\label{fig:19}
\end{figure} 

In the following section selected results of the repeated Latin Hypercube Sampling (rLHS) and the analysis of variance described above are shown. In all computations reported here, the average variance is computed as variance for each simulation day separately, and then averaged over the simulation days. 

The vertical level of highest variances in the updraft velocities (1200m above the ground) have been selected for visualization based on the vertical velocity variance time series plotted in the left panel of Figure~\ref{fig:19}. Based on a similar time series of the smoke concentrations and the plume top height estimates form the baseline case (Figure~\ref{fig:11}), the analysis level for smoke has been set to 1400m. Figure~\ref{fig:19}(b) shows the variance of the vertical wind velocity at 1200m above the terrain, computed from  25 rLHS-sampled simulations. Two well-defined maxima are located outside of the northwestern plot boundary, indicating optimal locations for upper-level vertical velocity measurements. Local variance reaches up to 3 $\mathrm{m}^2/\mathrm{s}^2$, which corresponds to local deviations in the vertical velocity of about 1.7 m/s between the sampled simulations, which is high enough to be sampled by a Doppler Lidar, Radar or Sodar. However, the fact that the updraft variances are concentrated over small regions indicate that optimal placement of the measurement devices may play a critical role in constraining model parameters based on the vertical velocity observations. 

\begin{figure}[tb]
\centering
\begin{tabular}{cc}
\includegraphics[width=7cm]{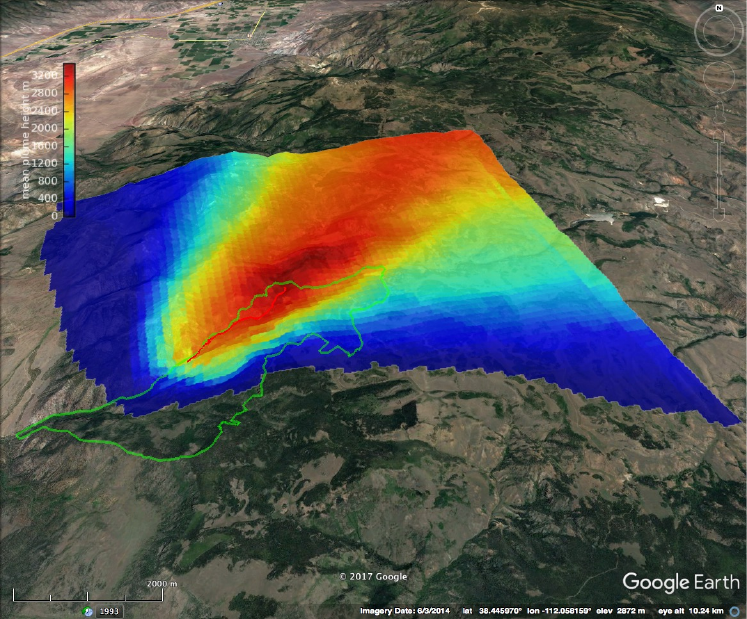}  &
\includegraphics[width=7cm]{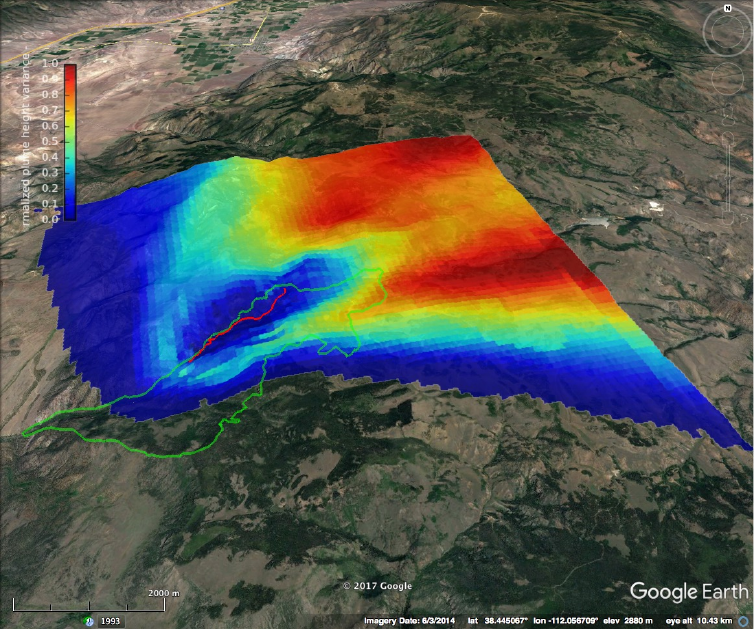} \\
(a) & (b) \\
\end{tabular} 
\caption{(\textbf{a}) The mean plume top height.  (\textbf{b}) The normalized variance in the plume height. The green contour represents the burning plot boundary and the red line shows the ignition.}
\label{fig:21}
\end{figure} 

\begin{figure}[tb]
\centering
\begin{tabular}{cc}
\multicolumn{2}{c}{\includegraphics[width=15cm]{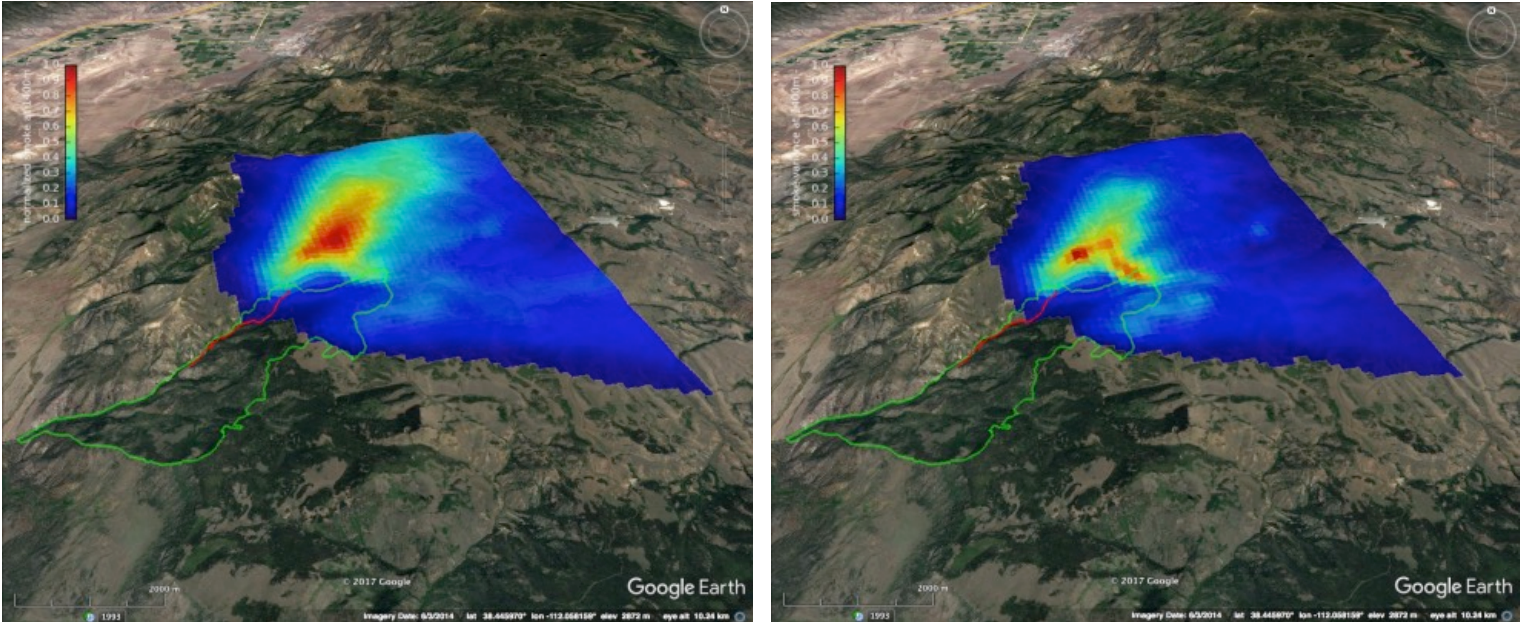}}\\
\makebox[7cm][c]{(a)} & \makebox[7cm][c]{(b)} \\
\end{tabular}
\caption{(\textbf{a}) The mean normalized smoke concentration at 1400m above the terrain. (\textbf{b}) The normalized smoke concentration average variance at 1400m above the terrain. The green contour represents the burning plot boundary, the red line shows the ignition.}
\label{fig:20}
\end{figure} 

Figure~\ref{fig:21} shows maps of the mean and the normalized variance of the plume top height. The former informs about the expected vertical plume extent which should be taken into account when the plume sampling strategies are considered. The latter shows regions of most pronounced impact of the model parameters on the plume rise. Interestingly, \emph{the region directly downwind from the fire, where the plume reaches highest elevations, is not the same as the region of highest plume height average variance}.  The plume height average variance exhibits high values over two symmetrical regions located on both sides of the plume axis and are pushed much more downwind from the fire compared to the mean plume top height. These results indicate that while the localized vertical smoke sampling (corkscrew path) may be optimal for assessing the maximum plume top height, the sampling region must be significantly extended downwind in order to adequately sample the plume top height variance. 

Figure~\ref{fig:20}(a) shows the map of the mean smoke concentration at 1400m, and Figure~\ref{fig:20}(b) is the map of the smoke concentration variance. The mean smoke concentration is intended to inform the measurements where the probability of successful smoke sampling is the highest. The variance on the other hand, shows where the tested model parameters have strongest impact on the smoke concentrations.  It is noteworthy that high values of smoke concentrations variances are found over much larger region than the vertical velocity variance shown in Figure~\ref{fig:19}(b), which suggest that vertical velocity measurements are more suitable for local platforms like ground Lidars, while the smoke measurements could be performed from airborne platforms covering larger areas. 

\begin{figure}[tb]
\centering
\begin{tabular}{cc}
\includegraphics[width=7cm]{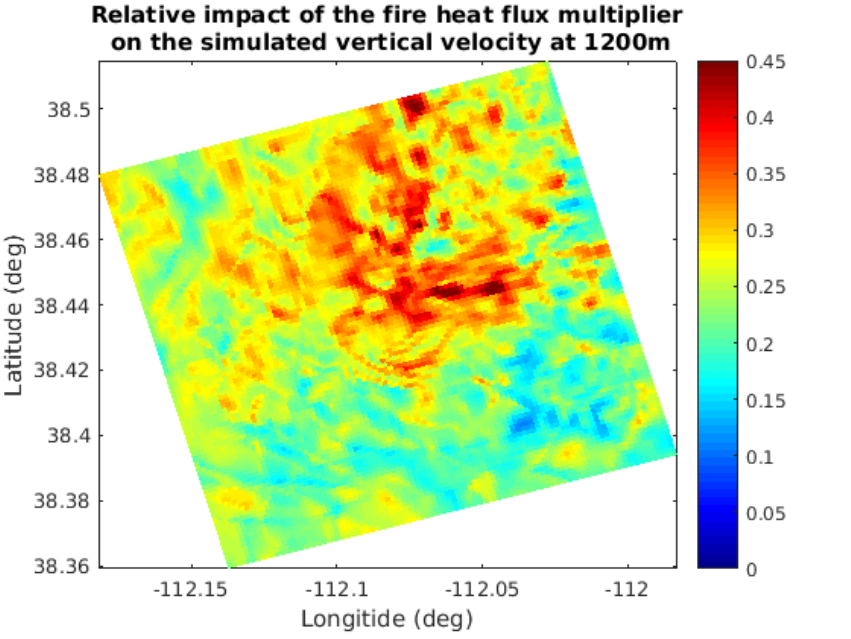}  &
\includegraphics[width=7cm]{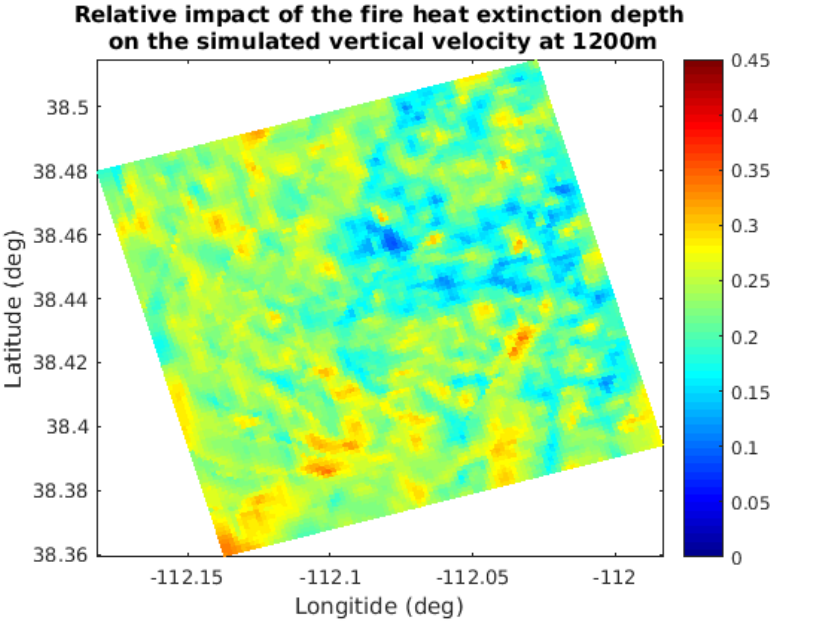} \\
\includegraphics[width=7cm]{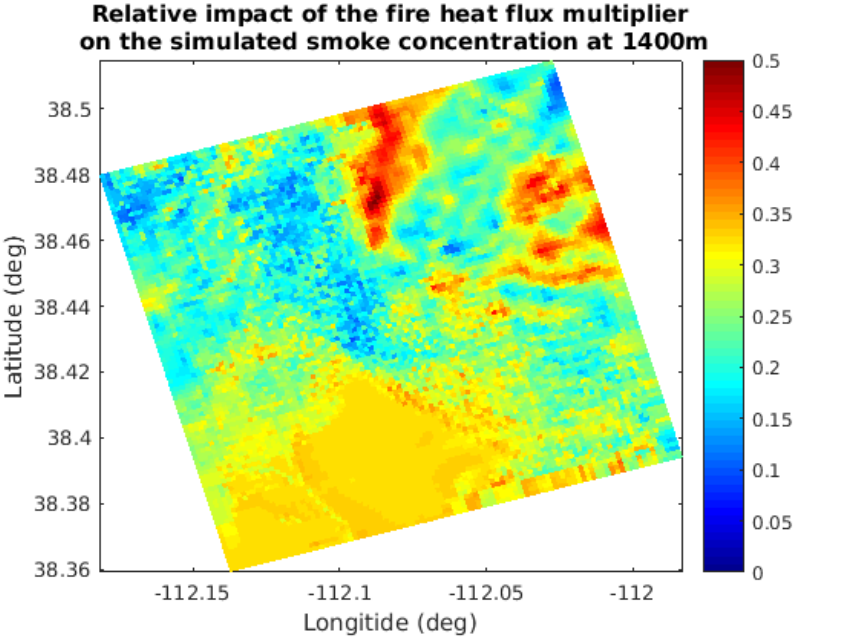}  &
\includegraphics[width=7cm]{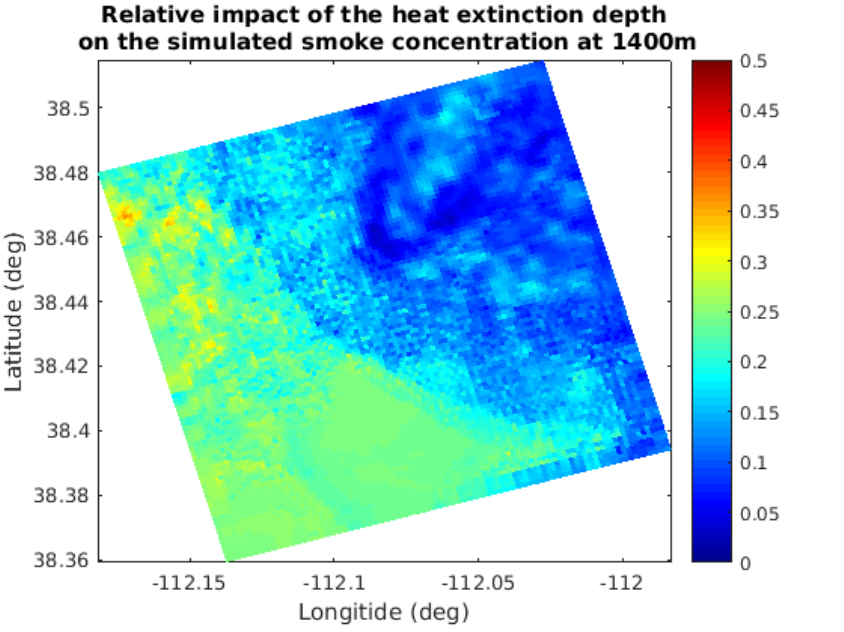} \\
\includegraphics[width=7cm]{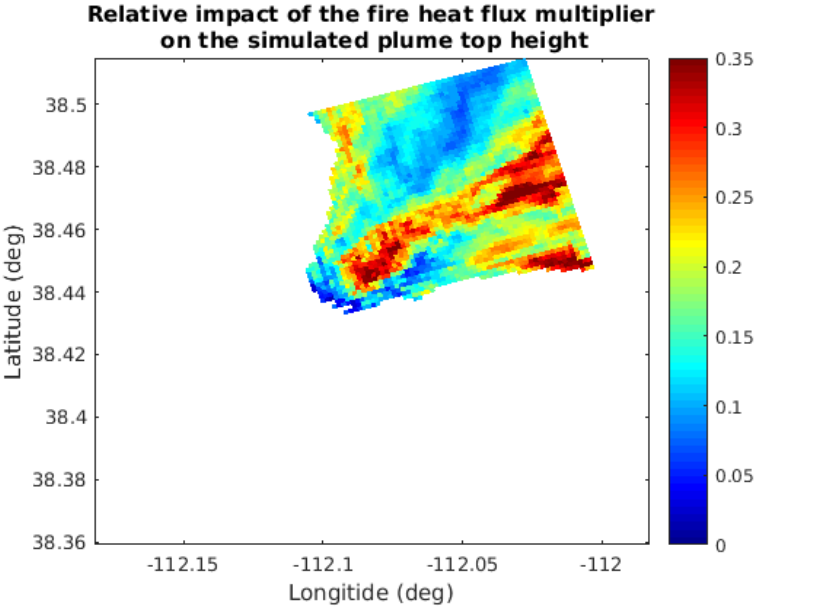}  &
\includegraphics[width=7cm]{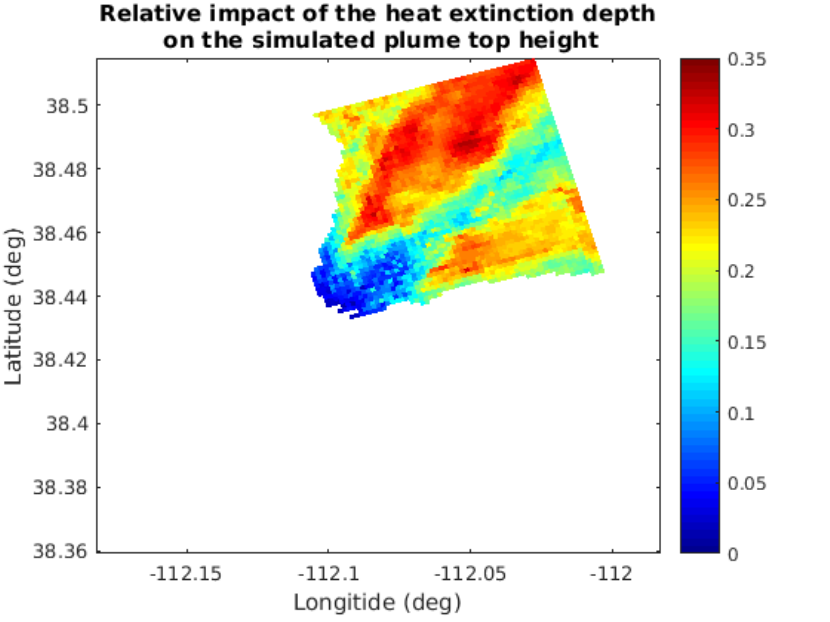} \\
\end{tabular} 
\caption{Sensitivity of the vertical velocity at 1200m (top), smoke concentration at 1400m (middle), and plume top height (bottom) to the fire heat flux multiplier (left column) and the heat extinction depth (right column).}
\label{fig:22}
\end{figure}

The rLHS analysis not only informs where the measurement should be taken in order to constrain model parameters, but also allows to find relative contribution of the tested parameters to variances in the variables of interest. 
We present results of the first-order variance decomposition of vertical velocity at 1200m, smoke concentration at 1400m, and the plume top height, attributed to the most important simulation parameters. 
These spatial plots of the sensitivity indices inform about the relative importance of the tested parameters (such as fuel moisture, heat extinction depth, fire heat flux, etc.) as well as indicate regions of highest impact of these parameters on the variables of interest. 

\afterpage{\clearpage}

The most critical parameters controlling the vertical velocity, smoke concentration and the plume top height are shown in Figure~\ref{fig:22}. The heat flux multiplier contributes to the variance of vertical velocity, smoke concentration and the plume top height up to 50\%. Unsurprisingly, the fire heat flux plays the major role in driving the plume rise, which indicates the importance of a detailed fire heat characterization during experimental burns. The second most significant contribution (up to 40\%) comes from the heat extinction depth, defining the depth over which the fire heat flux is distributed vertically in the model. The extinction depth seems to have an impact of similar magnitude on the variables of interest like the heat flux multiplier, but its character is different. Figure~\ref{fig:22} suggests that the heat flux multiplier has the opposite effect on the extinction depth. For instance, the vertical velocities are impacted by the heat flux multiplier mostly downwind (northeast) from the fire (see left top panel). The extinction depth, on the other hand, affects mostly vertical velocities upwind from the fire (see right top panel). Similarly, the smoke concentration is impacted by the heat flux in the north-eastern part of the domain, while the extinction depth seems to impact mostly the south-western part. A similar situation can be observed in the bottom panels of Figure~\ref{fig:22}, showing the impact of these parameters on the plume top height. Here, the heat flux impact is confined to the very core of the plume, while the extinction depth has the most pronounced impact on the sides of the plume. The differences in these patterns indicate that the placement of the sensor optimal form the perspective of constraining the heat flux may be not optimal for constraining the extinction depth and vice-versa. Also, the patterns of the heat flux contribution seem generally more organized than the ones of the extinction depth. The impact of the extinction depth seems to have non-local character especially when vertical velocity and smoke concentration is considered, spreading across large areas upwind from the fire. That suggests that estimating the extinction depth through the observations of vertical velocities and smoke concentration may be difficult, and may require direct tower-based observations of the vertical attenuation of the fire heat flux.

\begin{figure}[tb]
\centering
\begin{tabular}{cc}
\includegraphics[width=7cm]{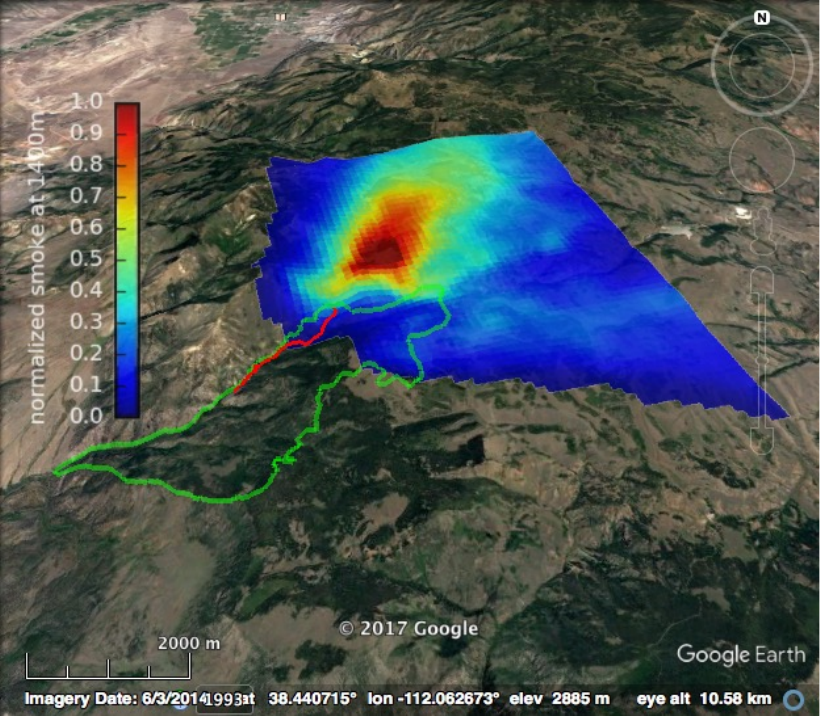}  &
\includegraphics[width=7cm]{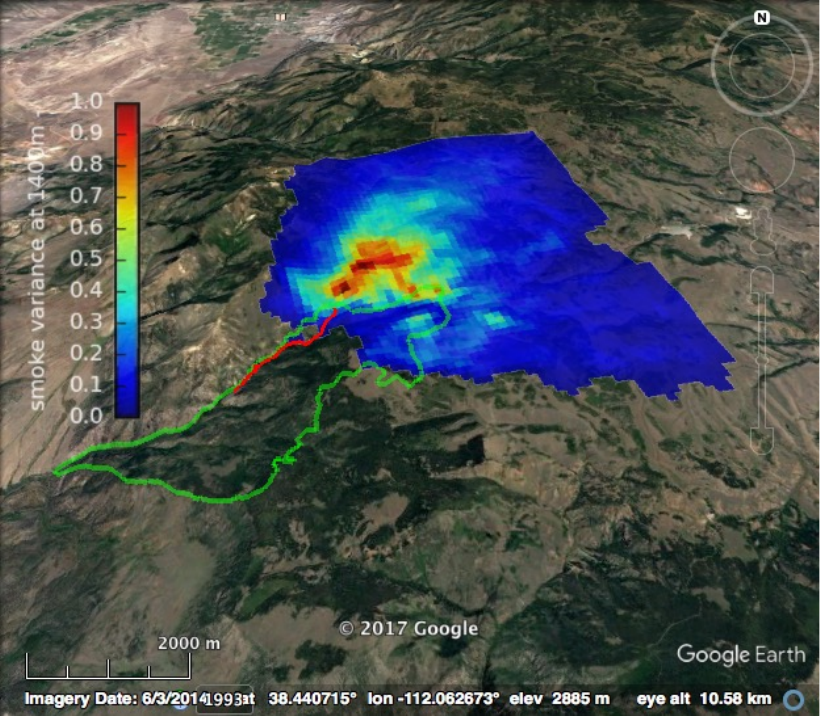} \\
(a) & (b) \\
\end{tabular} 
\caption{(\textbf{a}) The mean normalized smoke concentration at 1400m above the ground computed from 5 most typical days. (\textbf{b})  The normalized average variance of the smoke concentration at 1400m computed from 5 most typical days. Green contour represents burn unit boundaries, red line shows the ignition.}
\label{fig:23}
\end{figure} 

\begin{figure}[tb]
\centering
\begin{tabular}{cc}
\includegraphics[width=7cm]{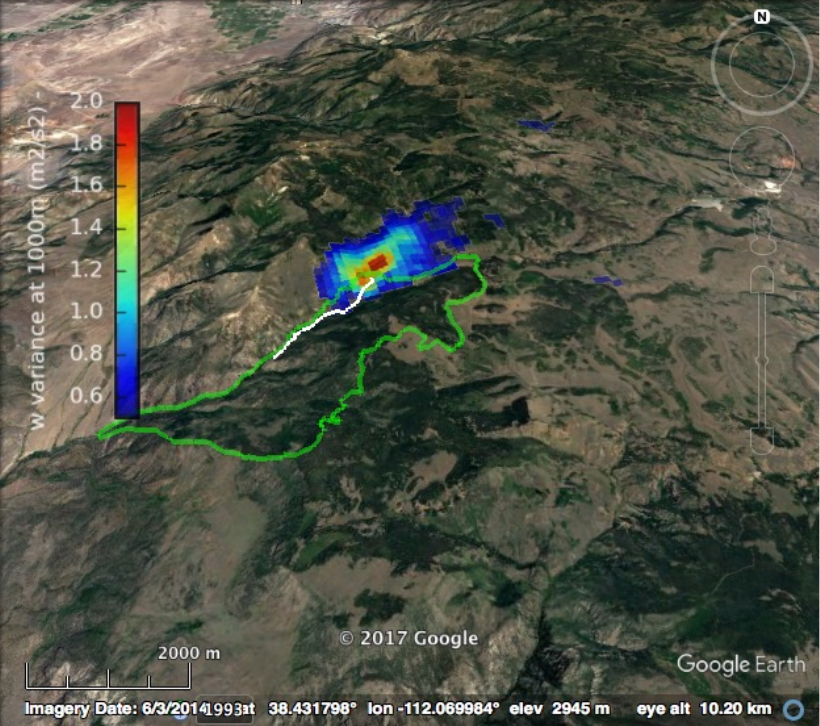}  &
\includegraphics[width=7cm]{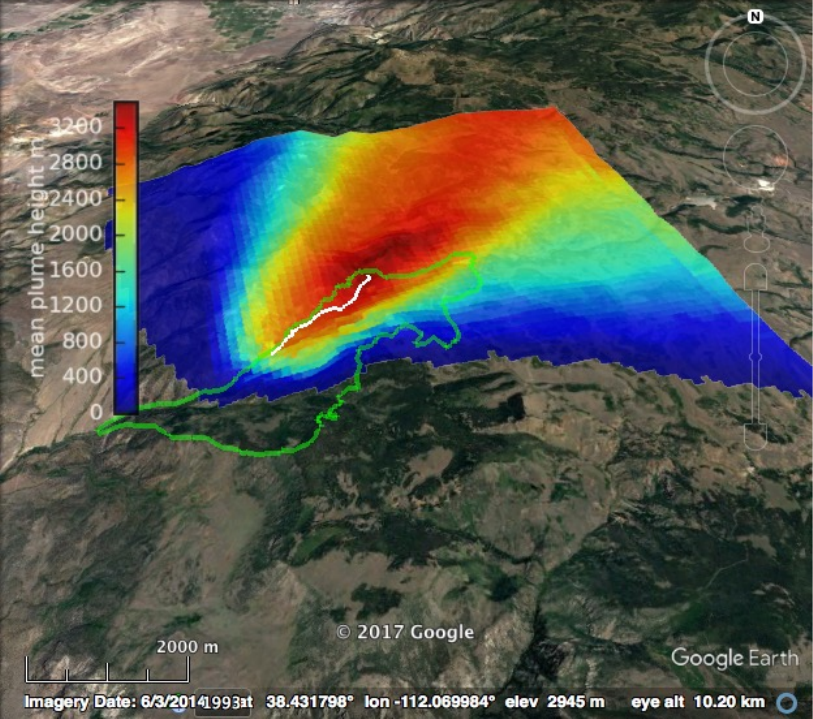} \\
(a) & (b) \\
\end{tabular} 
\caption{(\textbf{a}) The vertical velocity average variance at 1000m computed from 5 most typical days. (\textbf{b}) The mean plume top height computed from 5 most typical days. Green contour represents burn unit boundaries, white line shows the ignition.}
\label{fig:24}
\end{figure}

The analysis presented above focuses on the impact of model parameters on measurable variables like the vertical velocity or the plume height.
In that sense, it informs where the signal coming from the changes in the parameters is expected to be highest and consequently, shows locations where measurements should be taken to detect it. 
As the weather conditions on the burn day cannot be predicated at the moment of experimental planning, the analysis presented in this study has been employed to find most typical historical days for all the burn stations. 
Five most typical days have been included in this analysis. On the top of the baseline simulation, performed for 09.03.2014, analogous runs have been executed for other 4 typical days listed in Table~\ref{tab:typical}. 
These 5 days are treated as an ensemble of most typical scenarios for the Fishlake experimental burn. Shown below are the means and average variances of the variables of interest computed across the simulations performed for these days. Figure~\ref{fig:23} shows maps of mean and variance of the smoke concentration at 1400m. A pattern can be noticed there, indicating general north-northwest smoke dispersion. The fact that regions of maxima in both mean concentration and the variance can be easily identified, indicates that the typical days are similar to each other in terms of a general flow pattern in the Fishlake region. It should be noted that the wind direction was not a part of burn requirements for this burn site which defined only the wind speed, temperature, and relative humidity. Since typical days show relatively consistent flow pattern, it can be concluded that typical days in terms of the air humidity, wind speed, and moisture are generally associated with south-southwesterly flow. 

The maps of the vertical velocity variance and the mean plume top height are shown in Figure~\ref{fig:24}. A clear maximum in the updraft fluctuations is evident north-west from the end of the ignition line, as well as a well-defined north-northeasterly plume reaching 3200m. The statistical analysis of simulations performed for the most typical day such as the one shown in Figure~\ref{fig:23} and Figure~\ref{fig:24}, are intended to be used as a guideline in the planning stages of experimental burns when optimal locations of smoke and vertical velocity measurements are being considered.

\section{Conclusion}
\subsection{Estimation of typical days}
If the meteorological state vector $x$ had a multivariate Gaussian distribution, then the most typical day as defined in this work would also be the day with the most probable state. As Figure~\ref{fig:7} shows, a very typical day need not occur very often, or within an interval of similarly typical days; rather, it is a day when the state is in a rigorous sense very close to its sample mean. Indeed, for a non-Gaussian distribution (the case here), a day with state close to its mean need not occur very often, even approximately. From this standpoint, the analysis performed here assures that statistically, any historical day meeting the burn requirements will be similar to the identified typical day, which fits the purpose of this work. In future work, an alternative to consider would be to pick a state (and thus, the day it occurs) randomly from a multivariate bin containing states that do occur simultaneously very often.  However, before following this alternative, one would have to choose among many competing theories about how to choose the sizes of such a bin robustly. 

\subsection{Results from Numerical Experiments}
The numerical experiments presented in this study were performed and analyzed to aid with the preparation of the FASMEE experimental plan. Since we do not know how similar the typical burn days are to actual burn days, the results presented here cannot provide deterministic forecasts.  Thus, we can only provide examples of possible realizations of the planned experimental burns. A timely simulation (possibly with a truncated sensitivity study) can be run immediately before the burn to inform the deployment of observation resources, as it was successfully done for the FireFlux II experiment \cite{Clements-2014-OFG}. 

It has to be emphasized, that presented results of numerical simulations of anticipated burns should be treated as guidelines only. Even though WRF-SFIRE has been validated against vertical velocities measured during the FireFlux burn \cite{Kochanski-2013-EWP}, and the simulated plume top heights compared favorably with MISR observations \cite{Kochanski-2016-TIS}, its ability to realistically resolve the wildfire plume dynamics, and the near surface flow induced by a wildland fire remain to be assessed. Especially, in the context of simplified representation of the canopy flow in WRF, which may limit its ability to render near-surface winds in forested regions.

The results showing model sensitivity to grid resolution (not shown here) indicates that local updraft velocities may be significantly underestimated in the model simulations when the model mesh is not fine enough to resolve small-scale plume features. Therefore the simulated vertical velocities should be treated as lower limits for the expected values. The simulated updraft strengths suggest that measurement platforms used during the experiments should be robust enough to sustain updrafts exceeding 10 m/s. Our simulations suggest that the strongest updrafts will be observed in the Fishlake burn. However, due to the influence of complex local topography, exact identification of the fire impact on the local circulation may be difficult there. Therefore, the Fort Stewart burn performed in fat terrain would serve as a great complementary burn that could provide a clearer picture of fire-induced circulations. 

One of the most important aspects of experimental burns is their durations. As indicated by the western U.S. simulations, the burn experiments should last at least a couple of hours in order to fully capture the fire plume evolution, and consequently, the burn plots should be big enough to enable fire progression over this period of time. For small experimental burns, the fire may reach the end of the burn plot before the plume fully evolves. 

Unlike the fire-induced horizontal winds, expected to reach maximum speeds relatively close to the ground, maximum vertical velocities are expected at significantly higher levels. Simulations of the experimental burns indicate strong updrafts located up to a couple of kilometers aloft, so Doppler Lidars and other platforms taking vertical wind measurements should provide adequate vertical coverage.
Regarding smoke measurements, the simulations suggest that the vertical smoke concentration profiles and the plume top heights may vary significantly between the experimental burns. The Fishlake burn of expected highest intensity is associated with an elevated smoke concentration peak and higher plume tops than other lower-intensity burns. Therefore, it seems beneficial to adjust the measurement heights accordingly to assure optimal plume sampling.

Numerical simulations of the Fort Stewart burn indicate a strong sensitivity of the initial plume evolution to the fire ignition pattern. Therefore, from the standpoint of model validations, a detailed characterization of the ignition procedure is necessary. Also, as the fire initialization is crudely implemented in the model, interactions between simultaneously ignited fire lines are difficult to capture. Therefore, in experimental burns, simple ignitions are preferable, especially in cases when continuous IR observations at sufficient temporal and horizontal resolutions are not possible. Similarly, the sensitivity of the simulated plume top heights and vertical velocities to the fire heat flux (demonstrated based on the Fishlake simulations), indicates that detailed characterization of fire heat fluxes (radiative and convective) is critical to model evaluation. Convective heat fluxes from the fire must be observed during the experimental burns, in addition to radiative fluxes.

\subsection{Sensitivity Study of Sensor Placement}

The results presented in Section~\ref{sec:sampling} illustrate how numerical modeling combined with advanced statistical methods can objectively guide planning of experimental burns. There are multiple possible applications of this method. If an experiment is performed with the intention to constrain certain model parameters, such analysis can help assess the feasibility of this approach, and design an optimal sampling strategy. In the analyzed Fishlake burn, for instance, the results indicate that assessing the extinction depth through the vertical and smoke concentration measurements may not be feasible, which leads to a recommendation of direct measurements of the fire heat flux at various heights. This method also allows identification of the most critical model parameters that control plume dynamics. Here the heat flux has been identified as the one of highest importance for the plume development, which emphasizes the value of accurate characterization of fire heat fluxes during the experimental burns. As this method utilizes the results of statistical analysis of typical burn days it also informs about expected local flow patterns, plume dispersion and fire behavior in a probabilistic sense. If the decisions about sensor placements have to be made in advance, such analysis may guide them. In case of the Fishlake burn this analysis indicates that generally the area north-northeast from the burn plot should be considered for placing the measurement platforms.

We have performed only a pilot analysis of variations for several sampled variants of the Fishlake experimental burn. Listed below  are some preliminary observations and implications for future research:
\begin{itemize}[leftmargin=*,labelsep=4mm]
\item	The statistical analysis of runs with carefully sampled parameter sets was shown to provide clear guidelines on placing the measurements in space and time.
\item	Also, some measurements were found more affected by certain parameters, which can inform what parameters can be indirectly constrained by observations.
\item	Analyses of the variance in smoke concentration vertical velocity and plume top height from runs executed for different days and with different parameter sets show similar patterns indicating that typical day statistics managed to identify statistically similar days from the standpoint of plume rise and dispersion.
\item	The variability of the plume at higher altitudes (here, 1000-1400m above the ground) is concentrated up to couple kilometers downwind from the plot, which suggests that sampling right above the fire, optimal for the fire heat flux measurements, may not be optimal for sampling most active parts of the plume.
\item	Additional parameters can be considered. The cost of one repetition does not increase with the number of parameters, but more repetitions need to be done for statistical convergence. The variability of the outputs will increase and the added parameters will help model additional uncertainty, which is always present in reality.
\end{itemize}

\subsection{Computing Resources}
The rLHS computations reported here consisted of 1,000 simulations executed in about a week to reduce the sampling error. The wall clock time of one Fishlake simulation is about 2 hours on 3600 Intel cores, and it produces about 15GB of data. Counting repetition e.g., for runs failed for various reasons, for development, and for experimentations, about 1,000,000 core hours were used and 50TB of disk space was needed to store the outputs. After cleaning and elimination of redundant files and keeping only the final version, the final data set was 12TB. The statistical processing was done and the visualizations were generated on an auxiliary high-memory cluster in MATLAB. The analysis of the outputs of the simulations was performed in serial mode, which significantly contributed to the total analysis time. Overall, several thousands of hours on the auxiliary cluster were used.

As multiple simulations can be run concurrently, we suggest performing a similar analysis prior to the burn using available forecast data. With sufficient supercomputer allocation, this is feasible.

%%%%%%%%%%%%%%%%%%%%%%%%%%%%%%%%%%%%%%%%%%

%%%%%%%%%%%%%%%%%%%%%%%%%%%%%%%%%%%%%%%%%%
%\vspace{6pt} 

%%%%%%%%%%%%%%%%%%%%%%%%%%%%%%%%%%%%%%%%%%
%% optional
\supplementary{The WRF-SFIRE and WRFx software (wrfxpy, wrfxctrl, wrxfweb) are available at \url{https://github.com/openwfm}. Selected visualizations can be accessed at \url{http://demo.openwfm.org/fasmee}. MATLAB software developed for the experimental design is available at \url{https://github.com/openwfm/design}.
}

%%%%%%%%%%%%%%%%%%%%%%%%%%%%%%%%%%%%%%%%%%
\acknowledgments{This research was sponsored in part by the Joint Fire Science Program under grant JFSP 16-4-5-03 as a part of the Fire and Smoke Model Evaluation Experiment (FASMEE) Phase I. The authors would also like to acknowledge support of the National Science Foundation grants DMS-1216481, ICER-1664175, the NASA grant NNX13AH59G to Colorado State University, Sher Schranz, PI, CIRA in affiliation with NOAA/ESRL. 

The authors would like to acknowledge high-performance computing support from Yellowstone (ark:/85065/d7wd3xhc) and Cheyenne (doi:10.5065/D6RX99HX) provided by NCAR's Computational and Information Systems Laboratory, sponsored by the National Science Foundation. The computing support from the University of Utah Center for High Performance Computing and the Center for Computational Mathematics, University of Colorado Denver, is greatly appreciated. 

The authors would like to thank Derek Mallia for reading the paper and suggesting improvements, and anonymous reviewers for helpful comments, which resulted in improving the paper.}

%%%%%%%%%%%%%%%%%%%%%%%%%%%%%%%%%%%%%%%%%%
\authorcontributions{A.K. set up and analyzed the computational experiments and wrote the introduction, the computational experiments section,  and the conclusion. A.F. devised the statistical analysis of typical burn days, wrote the statistical analysis section, and suggested the use of rLHS and Sobol variance decomposition for the sensitivity analysis. J.M. made the custom developments within the WRF-SFIRE and WRFx softwares to support this work, devised and performed the sensitivity analysis, wrote the sensitivity analysis section, and finalized the paper for submission.}

%%%%%%%%%%%%%%%%%%%%%%%%%%%%%%%%%%%%%%%%%%
\conflictofinterests{The authors declare no conflict of interest. The funding sponsors had no role in the design of the study; in the collection, analyses, or interpretation of data; in the writing of the manuscript, and in the decision to publish the results.} 

%%%%%%%%%%%%%%%%%%%%%%%%%%%%%%%%%%%%%%%%%%
%% optional
%\abbreviations{The following abbreviations are used in this manuscript:\\

%\noindent MDPI: Multidisciplinary Digital Publishing Institute\\
%DOAJ: Directory of open access journals\\
%TLA: Three letter acronym\\
%LD: linear dichroism}

%%%%%%%%%%%%%%%%%%%%%%%%%%%%%%%%%%%%%%%%%%
%% optional
\appendix

\section{Statistics of Weather Data}
\label{app:weather}

This appendix contains a visualization of the remaining weather station data to complement Figures~\ref{fig:5} and~\ref{fig:6} in the description of the method in Section~\ref{sec:typical}.

\begin{figure}[tbph]
\centering
\includegraphics[width=14cm]{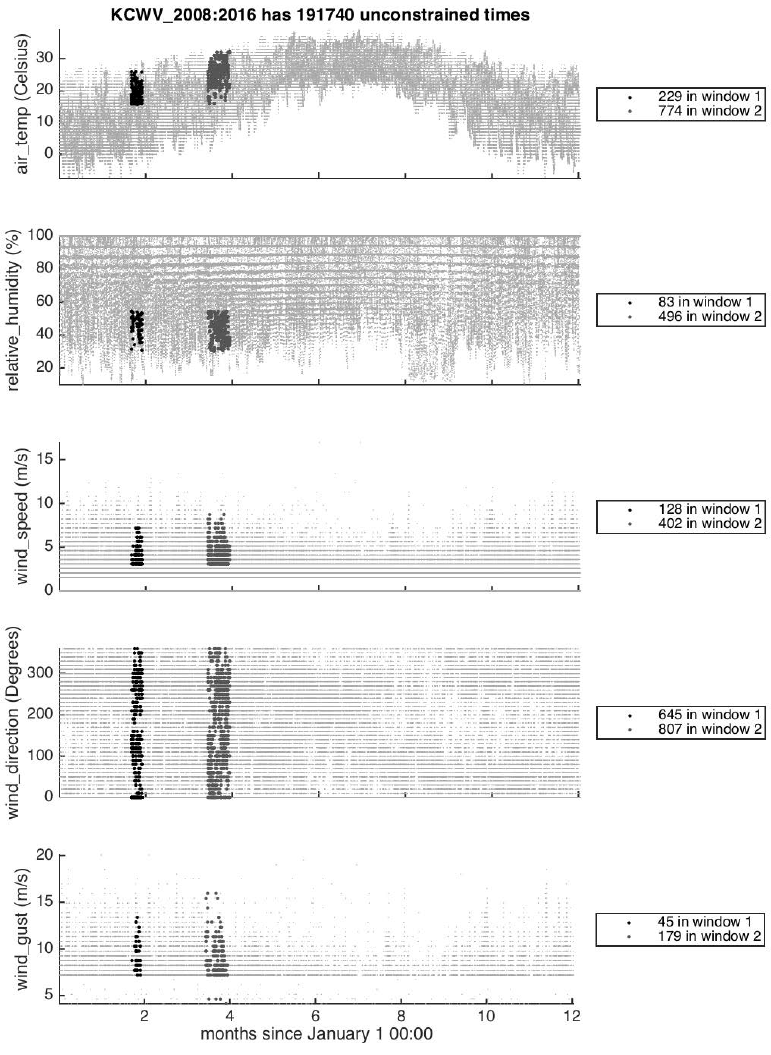}
\caption{Station data (vertical) vs month number (horizontal) for station KCWV (Claxton Evans County Airport, 32.19510\degree N, 81.86960\degree W 111 ft in Georgia) from 2008 to 2016 with 2 burn windows per year. Points that do and do not meet the burn-requirements (Table~\ref{tab:req}) are dark and light gray, respectively, and enumerated in the legends and titles, respectively.
The abscissa is $t\pmod{12}$ for all the years indicated in the title.}
\label{fig:25}
\end{figure} 

\begin{figure}[tbph]
\centering
\includegraphics[width=14cm]{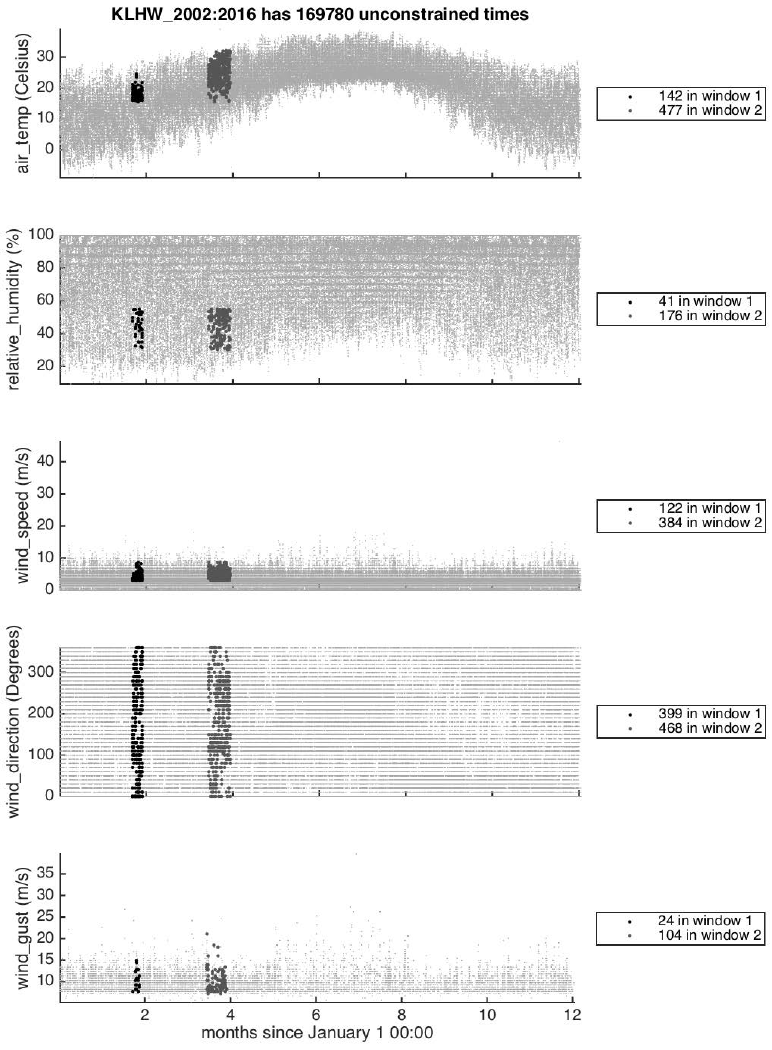}
\caption{Station data (vertical) vs month number (horizontal) for station KLHW (Ft. Stewart, 31.88333\degree N, 81.56667\degree W, 46 ft in Georgia) from 2002 to 2016 with 2 burn windows per year. Points that do and do not meet the burn-requirements (Table~\ref{tab:req}) are dark and light gray, respectively, and enumerated in the legends and titles, respectively.
The abscissa is $t\pmod{12}$ for all the years indicated in the title.}
\label{fig:26}
\end{figure} 

\begin{figure}[tbph]
\centering
\includegraphics[width=14cm]{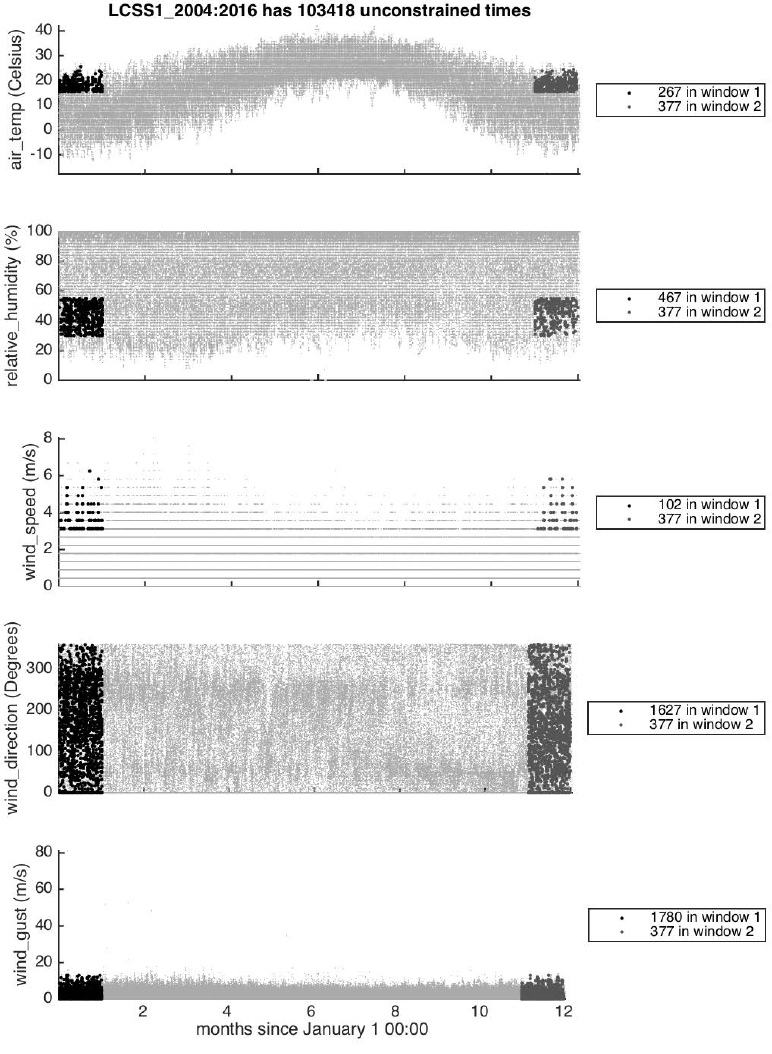}
\caption{Station data (vertical) vs month number (horizontal) for station LCSS1 (Savriv, 33.33305\degree N, 81.591667\degree W, 275 ft in South Carolina) from 2004 to 2016. Points that do and do not meet the burn-requirements (Table~\ref{tab:req}) are dark and light gray, respectively, and enumerated in the legends and titles, respectively.
The abscissa is $t\pmod{12}$ for all the years indicated in the title.}
\label{fig:27}
\end{figure} 

\begin{figure}[tbph]
\centering
\includegraphics[width=14cm]{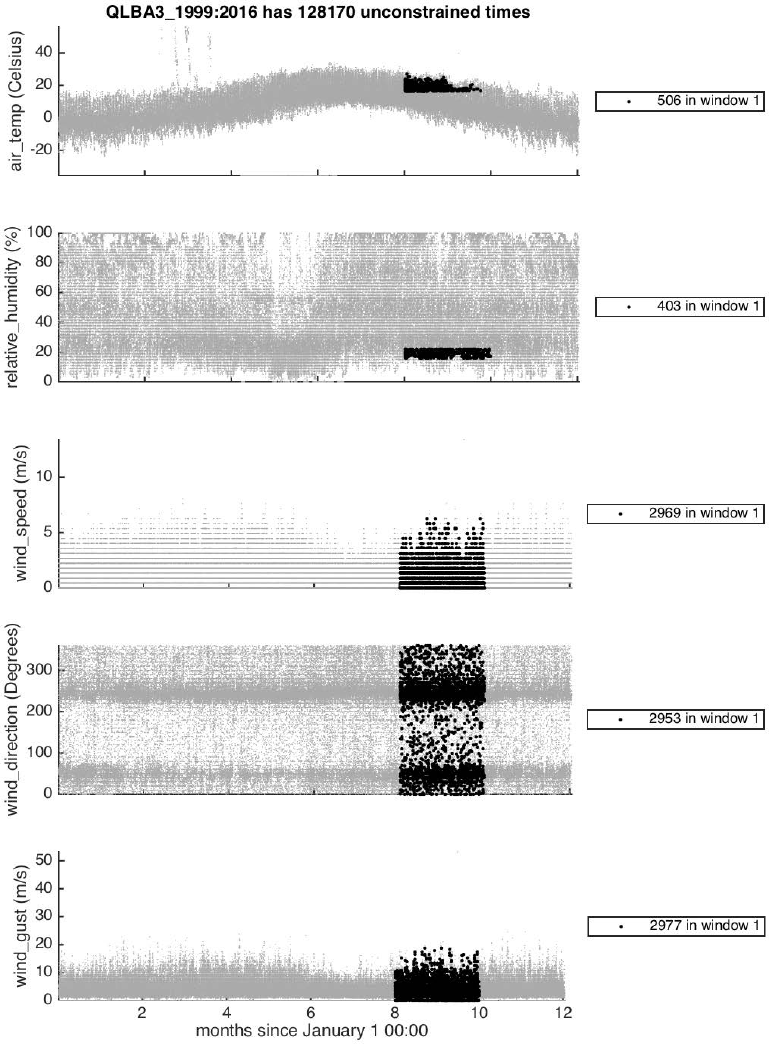}
\caption{Station data (vertical) vs month number (horizontal) for station QLBA3 (Lindbergh Hill, 36.285556\degree N, 112.078611\degree W, 8800 ft in Arizona) for 1999--2016. Points that do and do not meet the burn-requirements (Table~\ref{tab:req}) are dark and light gray, respectively, and enumerated in the legends and titles, respectively.
The abscissa is $t\pmod{12}$ for all the years indicated in the title.}
\label{fig:28}
\end{figure} 

\begin{figure}[tbph]
\centering
\includegraphics[width=14cm]{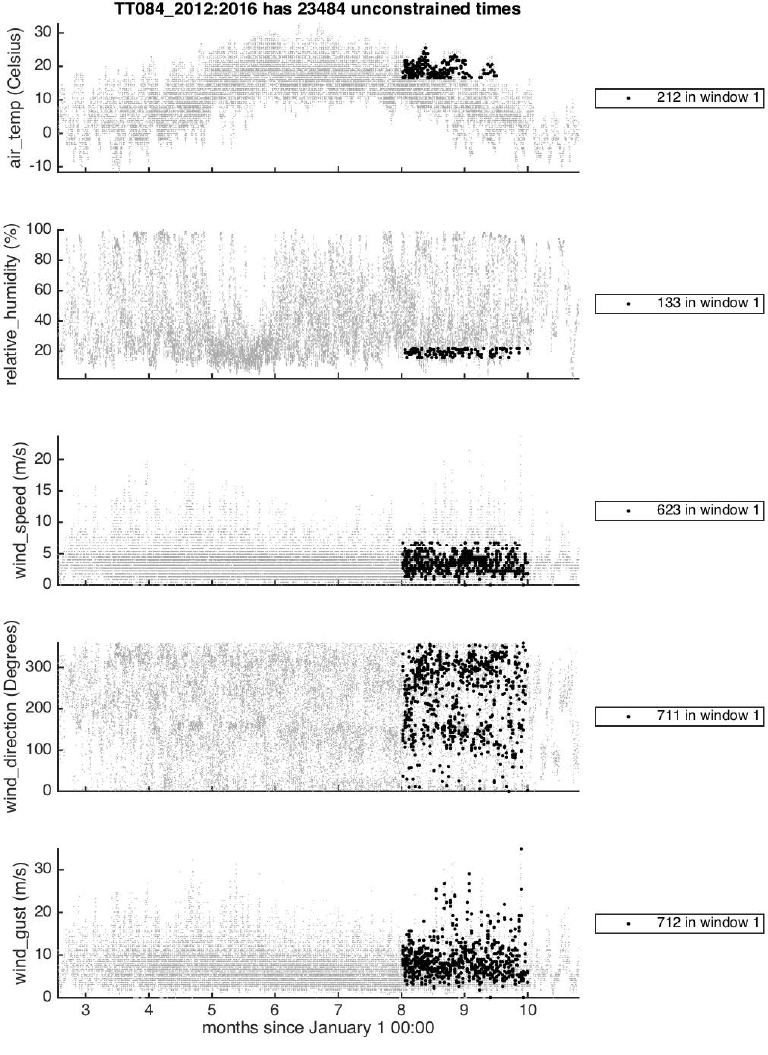}
\caption{Station data (vertical) vs month number (horizontal) for station TT084 (Fishlake D4 Pt \#4, 38.960319\degree N, 111.405983\degree W, 8523 ft in Utah) for 2012--2016. Points that do and do not meet the burn-requirements (Table~\ref{tab:req}) are dark and light gray, respectively, and enumerated in the legends and titles, respectively.
The abscissa is $t\pmod{12}$ for all the years indicated in the title.}
\label{fig:29}
\end{figure} 

\begin{figure}[H]
\centering
\includegraphics[width=12cm]{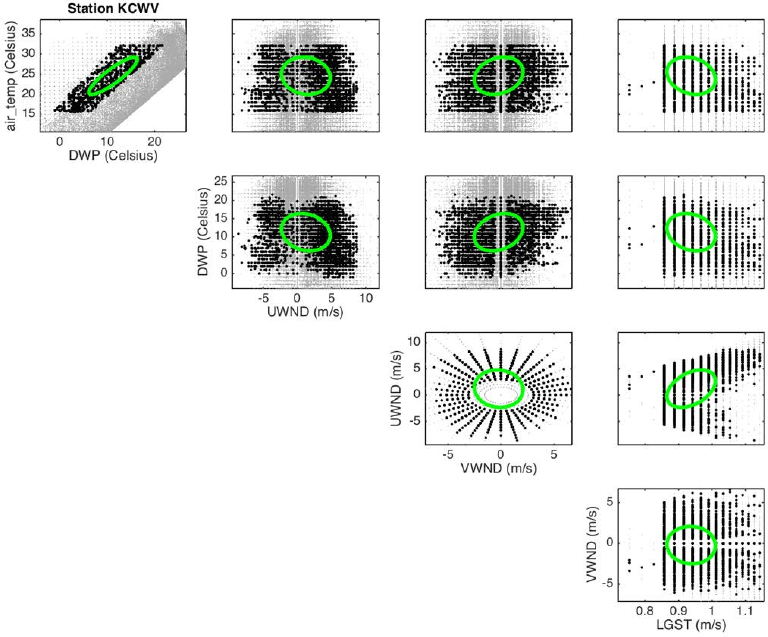}
\caption{As in Figure~\ref{fig:6} but for station KCVW.}
\label{fig:30}
\end{figure} 

\begin{figure}[H]
\centering
\includegraphics[width=12cm]{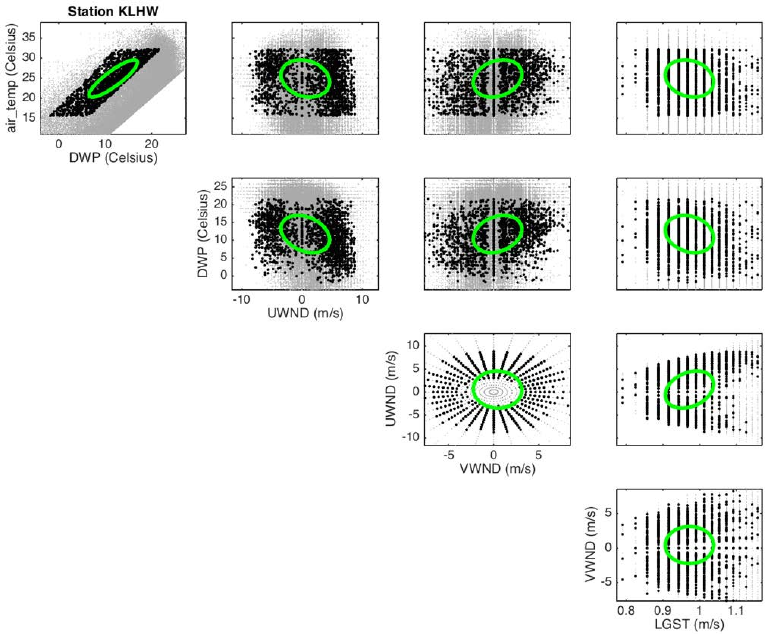}
\caption{As in Figure~\ref{fig:6} but for station KLHW.}
\label{fig:31}
\end{figure} 

\begin{figure}[H]
\centering
\includegraphics[width=12cm]{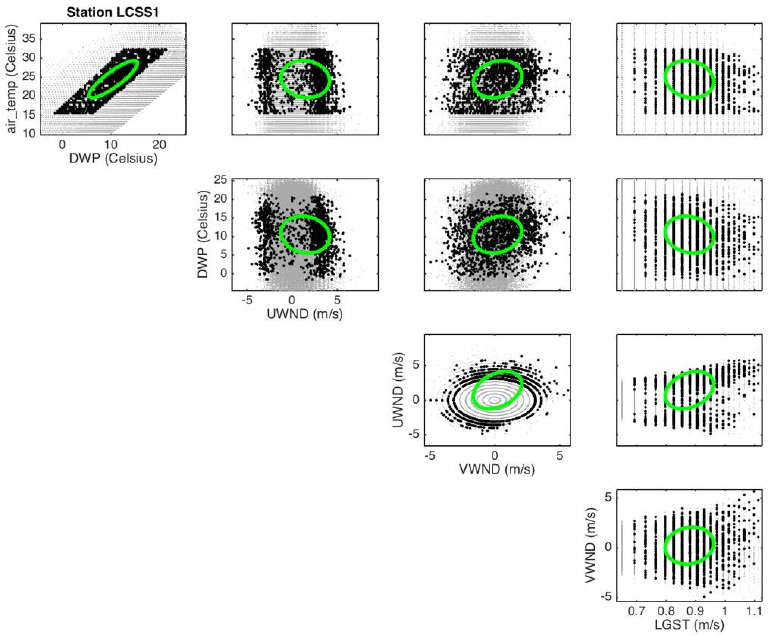}
\caption{As in Figure~\ref{fig:6} but for station LCSS1.}
\label{fig:32}
\end{figure} 

\begin{figure}[H]
\centering
\includegraphics[width=12cm]{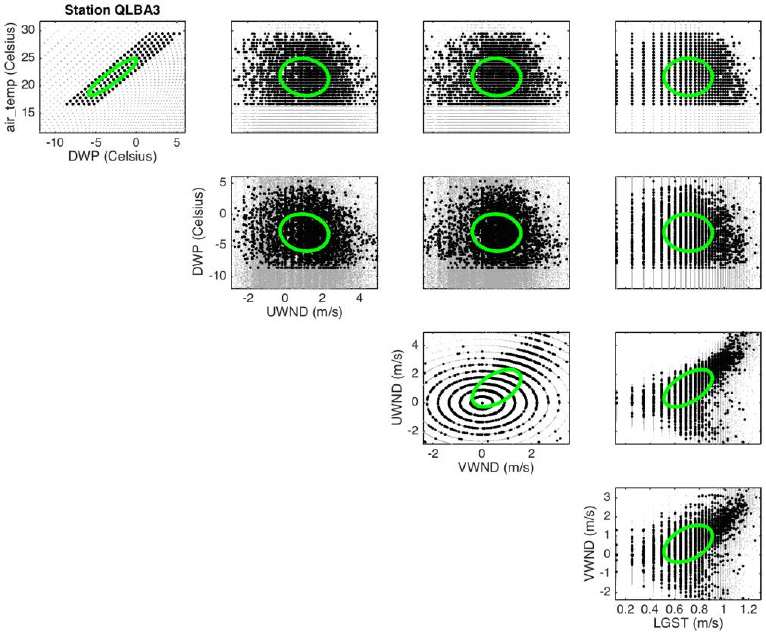}
\caption{As in Figure~\ref{fig:6} but for station QLBA3.}
\label{fig:33}
\end{figure} 

\begin{figure}[H]
\centering
\includegraphics[width=12cm]{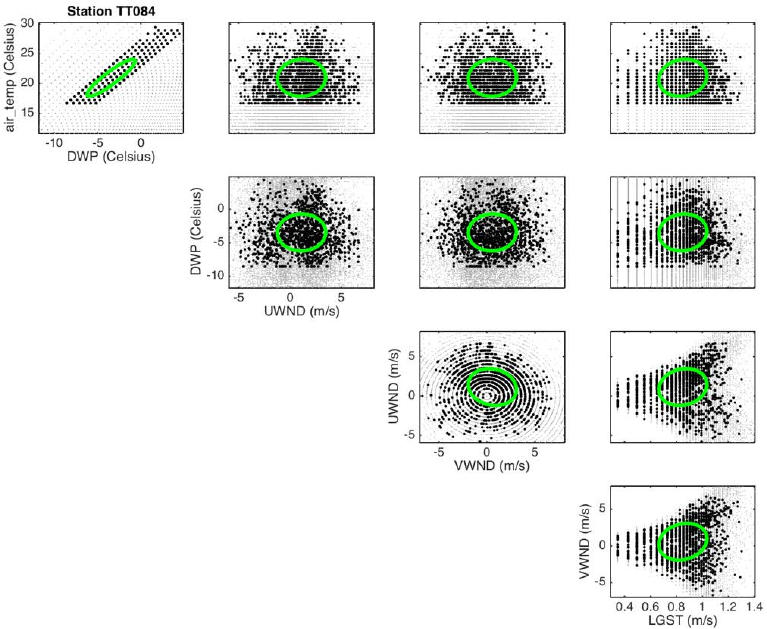}
\caption{As in Figure~\ref{fig:6} but for station TT084.}
\label{fig:34}
\end{figure}

%%%%%%%%%%%%%%%%%%%%%%%%%%%%%%%%%%%%%%%%%%
%\bibliographystyle{mdpi}

%=====================================
% References, variant A: internal bibliography
%=====================================
%\renewcommand\bibname{References}
%\begin{thebibliography}{999}
% Reference 1
%\bibitem{ref-journal}
%Lastname, F.; Author, T. The title of the cited article. {\em Journal Abbreviation} {\bf 2008}, {\em 10}, 142-149.
% Reference 2
%\bibitem{ref-book}
%Lastname, F.F.; Author, T. The title of the cited contribution. In {\em The Book Title}; Editor, F., Meditor, A., Eds.; Publishing House: City, Country, 2007; pp. 32-58.
%\end{thebibliography}

%=====================================
% References, variant B: external bibliography
%=====================================
\section*{References}
\bibliography{fasmee,other,geo,slides}

%%%%%%%%%%%%%%%%%%%%%%%%%%%%%%%%%%%%%%%%%%
%% optional
%\sampleavailability{Samples of the compounds ...... are available from the authors.}

%%%%%%%%%%%%%%%%%%%%%%%%%%%%%%%%%%%%%%%%%%
\end{document}